%% file: main.tex
\documentclass[preprint,12pt,authoryear]{elsarticle}
\input{local}

\input{mycommand}
\usepackage{amssymb}
\usepackage{amsmath}
\journal{Journal of Systems and Software}

\begin{document}

\begin{frontmatter}



\title{An Empirical Study of CGO Usage in Go Projects - Distribution, Purposes, Patterns and Critical Issues}


\author[cs]{Jinbao Chen}
\ead{zkd18cjb@mail.ustc.edu.cn}
\author[cs]{Boyao Ding}
\ead{via@mail.ustc.edu.cn}
\author[cs]{Yu Zhang\corref{cor1}}
\ead{yuzhang@ustc.edu.cn}
\cortext[cor1]{Corresponding author}
\author[cs]{Qingwei Li}
\ead{lqw332664203@mail.ustc.edu.cn}
\author[cs]{Fugen Tang}
\ead{fgtang@mail.ustc.edu.cn}
\affiliation[cs]{organization={School of Computer Science and Technology, University of Science and Technology of China},%Department and Organization
            addressline={100 Fuxing Road}, 
            city={Hefei},
            postcode={230031}, 
            state={Anhui},
            country={China}}

\input{abstract}

\end{frontmatter}


\input{intro}

\input{related}

\input{methodology}

\input{rq}

\input{ptrchk}

\input{proposal}
\input{threats}
\input{conclusion}
\input{data}
\input{ack}
\bibliography{go}
\bibliographystyle{elsarticle-harv} 
\end{document}

%% file: local.tex
\usepackage{graphicx}
\usepackage{booktabs}
\usepackage{longtable}
\usepackage[ruled,linesnumbered]{algorithm2e}
\usepackage{pifont}
\usepackage{multirow}
\usepackage{xspace}
\usepackage{url}
\usepackage{listings}
\usepackage{verbatim}
\usepackage{framed}
\usepackage{graphicx}  %插入图片的宏包
\usepackage{float}  %设置图片浮动位置的宏包
\usepackage{subfig}  %插入多图时用子图显示的宏包
\usepackage{fancyhdr}
\usepackage{xcolor}
\usepackage{ifthen}
\usepackage{caption}
\usepackage{soul}	
\usepackage{hyperref}

\lstdefinestyle{Go}{
    language=Go,
    basicstyle=\scriptsize\ttfamily, % 使用等宽字体
    keywordstyle=\color{blue}, % Go 关键字颜色为蓝色
    commentstyle=\color{green!60!black}, % 注释颜色为绿色
    stringstyle=\color{red}, % 字符串颜色为红色
    numbers=left, % 行号位置在左侧
    numberstyle=\tiny\color{gray}, % 设置行号字号为 tiny，颜色为灰色
    stepnumber=1, % 每行显示行号
    tabsize=1, % Tab 键宽度为 4 个空格
    showspaces=false, % 不显示空格
    showstringspaces=false, % 不显示字符串中的空格
    breaklines=true, % 自动换行
    frame=tb, % 代码块周围加框
    framexleftmargin=0mm, % 代码块框距离左侧页面边缘的距离
    framexrightmargin=0mm, % 代码块框距离右侧页面边缘的距离
    xleftmargin=0mm, % 代码块内容距离左侧框线的距离
    xrightmargin=0mm, % 代码块内容距离右侧框线的距离
    backgroundcolor=\color{gray!10}, % 代码块背景颜色
    captionpos=b % 标题位置在底部
}

\lstdefinestyle{Gotiny}{
    language=Go,
    basicstyle=\tiny\ttfamily, % 使用等宽字体
    keywordstyle=\color{blue}, % Go 关键字颜色为蓝色
    commentstyle=\color{green!60!black}, % 注释颜色为绿色
    stringstyle=\color{red}, % 字符串颜色为红色
    numbers=left, % 行号位置在左侧
    numberstyle=\tiny\color{gray}, % 设置行号字号为 tiny，颜色为灰色
    stepnumber=1, % 每行显示行号
    tabsize=1, % Tab 键宽度为 4 个空格
    showspaces=false, % 不显示空格
    showstringspaces=false, % 不显示字符串中的空格
    breaklines=true, % 自动换行
    frame=tb, % 代码块周围加框
    framexleftmargin=0mm, % 代码块框距离左侧页面边缘的距离
    framexrightmargin=0mm, % 代码块框距离右侧页面边缘的距离
    xleftmargin=0mm, % 代码块内容距离左侧框线的距离
    xrightmargin=0mm, % 代码块内容距离右侧框线的距离
    backgroundcolor=\color{gray!10}, % 代码块背景颜色
    captionpos=b % 标题位置在底部
}

\lstdefinestyle{Cpp}{
    language=C++, % 指定语言为 C++
    basicstyle=\scriptsize\ttfamily, 
    keywordstyle=\color{blue}, % C++ 关键字颜色为蓝色
    commentstyle=\color{green!60!black}, % 注释颜色为绿色
    stringstyle=\color{red}, % 字符串颜色为红色
    numbers=left, % 行号位置在左侧
    numberstyle=\tiny\color{gray}, % 设置行号字号为 tiny，颜色为灰色
    stepnumber=1, % 每行显示行号
    tabsize=1, % Tab 键宽度为 4 个空格
    showspaces=false, % 不显示空格
    showstringspaces=false, % 不显示字符串中的空格
    breaklines=true, % 自动换行
    frame=tb, % 代码块周围加框
    framexleftmargin=0mm, % 代码块框距离左侧页面边缘的距离
    framexrightmargin=0mm, % 代码块框距离右侧页面边缘的距离
    xleftmargin=0mm, % 代码块内容距离左侧框线的距离
    xrightmargin=0mm, % 代码块内容距离右侧框线的距离
    backgroundcolor=\color{gray!10}, % 代码块背景颜色
    captionpos=b % 标题位置在底部
}

\lstdefinestyle{Shell}{
    language=bash, % 指定语言为 Bash
    basicstyle=\scriptsize\ttfamily, 
    keywordstyle=\color{blue}, % Bash 关键字颜色为蓝色
    commentstyle=\color{green!60!black}, % 注释颜色为绿色
    stringstyle=\color{red}, % 字符串颜色为红色
    numbers=left, % 行号位置在左侧
    numberstyle=\tiny\color{gray}, % 设置行号字号为 tiny，颜色为灰色
    stepnumber=1, % 每行显示行号
    tabsize=1, % Tab 键宽度为 4 个空格
    showspaces=false, % 不显示空格
    showstringspaces=false, % 不显示字符串中的空格
    breaklines=true, % 自动换行
    frame=tb, % 代码块周围加框
    framexleftmargin=0mm, % 代码块框距离左侧页面边缘的距离
    framexrightmargin=0mm, % 代码块框距离右侧页面边缘的距离
    xleftmargin=0mm, % 代码块内容距离左侧框线的距离
    xrightmargin=0mm, % 代码块内容距离右侧框线的距离
    backgroundcolor=\color{gray!10}, % 代码块背景颜色
    captionpos=b % 标题位置在底部
}

\setstcolor{red}

\long\def\com#1{}
\newboolean{showremv}
\setboolean{showremv}{false} % set to true if you want to show removed text

\ifthenelse{\boolean{showremv}}
{
\newcommand{\remv}[1]{{\color{magenta}\st{#1}}}
}
{
\newcommand{\remv}[1]{{}}
}

\ifthenelse{\boolean{showremv}}
{

\newcommand{\cjb}[1]{\textcolor{blue}{#1}}
\newcommand{\cjbr}[2]{%
  \remv{#1}%    先删除
  \ \cjb{#2}%        再标出新内容
}
}
{

\newcommand{\cjb}[1]{{#1}}
\newcommand{\cjbr}[2]{{#2}}
}

\newcommand{\mysect}[1]{\textbf{\textit{#1}}}		% subsubsection
\newcommand{\kw}[1]{{\it #1}}	% keyword
\newcommand{\kn}[1]{\texttt{\small #1}}	% keyword
\newcommand{\kd}[1]{\textbf{#1}}	% keyword
\newcommand{\codet}[1]{{\small \texttt{#1}}}	% small textsf

\newcommand\eg{{\it{e.g.,\ }}}

\newcommand{\eat}[1]{\if 0 #1 \fi}

%% file: mycommand.tex
\usepackage{framed}
\usepackage{ulem}
\newcommand{\finding}[2]{\setlength{\fboxrule}{0.2pt}\begin{framed}
\noindent \textbf{Finding #1:} #2
\end{framed}}

\newcommand{\mysys}{\textit{CGOAnalyzer}\xspace}
\newcommand{\myinsert}[1] {#1}
\newcommand{\mydel}[1] {}

\newcommand{\myins}[1]{#1}
\newcommand{\del}[1]{}

%% file: abstract.tex
\begin{abstract}
%\cjbnote{no or short story in abstract.}
Multilingual software development integrates multiple languages into a single application, with the Foreign Function Interface (FFI) enabling seamless interaction. While FFI boosts efficiency and extensibility, it also introduces risks. 
%The adoption of FFI not only significantly enhances development efficiency but also elevates the scalability of the software.
%While traditional high-level languages like Python and Java provide mature cross-language interfaces (e.g., Python/C API, JNI), Go, as an emerging language, offers CGO for this purpose. Although multi-language development brings several advantages, it also introduces challenges due to differences in runtime systems and memory management among languages. 
%Understanding the usage of FFI is crucial for improving development efficiency and reducing potential risks. However, 
Existing studies focus on FFIs in languages like Python and Java, neglecting CGO, the emerging FFI in Go, which poses unique risks. 

To address these concerns, we conduct an empirical study of CGO usage across 920 open-source Go projects. Our study aims to reveal the distribution, patterns, purposes, and critical issues associated with CGO, offering insights for developers and the Go team. We develop \mysys, a tool to efficiently identify and quantify CGO-related features. Our findings reveal that: (1) 11.3\% of analyzed Go projects utilize CGO, with usage concentrated in a subset of projects; (2) CGO serves 4 primary purposes, including system-level interactions and performance optimizations, with 15 distinct usage patterns observed; (3) 19 types of CGO-related issues exist, including one critical issue involving unnecessary pointer checks that pose risks of runtime crashes due to limitations in the current Go compilation toolchain; (4) a temporary solution reduces unnecessary pointer checks, mitigating crash risks, and (5) we submitted a proposal to improve the Go toolchain for a permanent fix, which has been grouped within an accepted proposal for future resolution. Our findings provide valuable insights for developers and the Go team, enhancing development efficiency and reliability while improving the robustness of the Go toolchain.

\end{abstract}

\begin{graphicalabstract}
\includegraphics[width=1.0\textwidth]{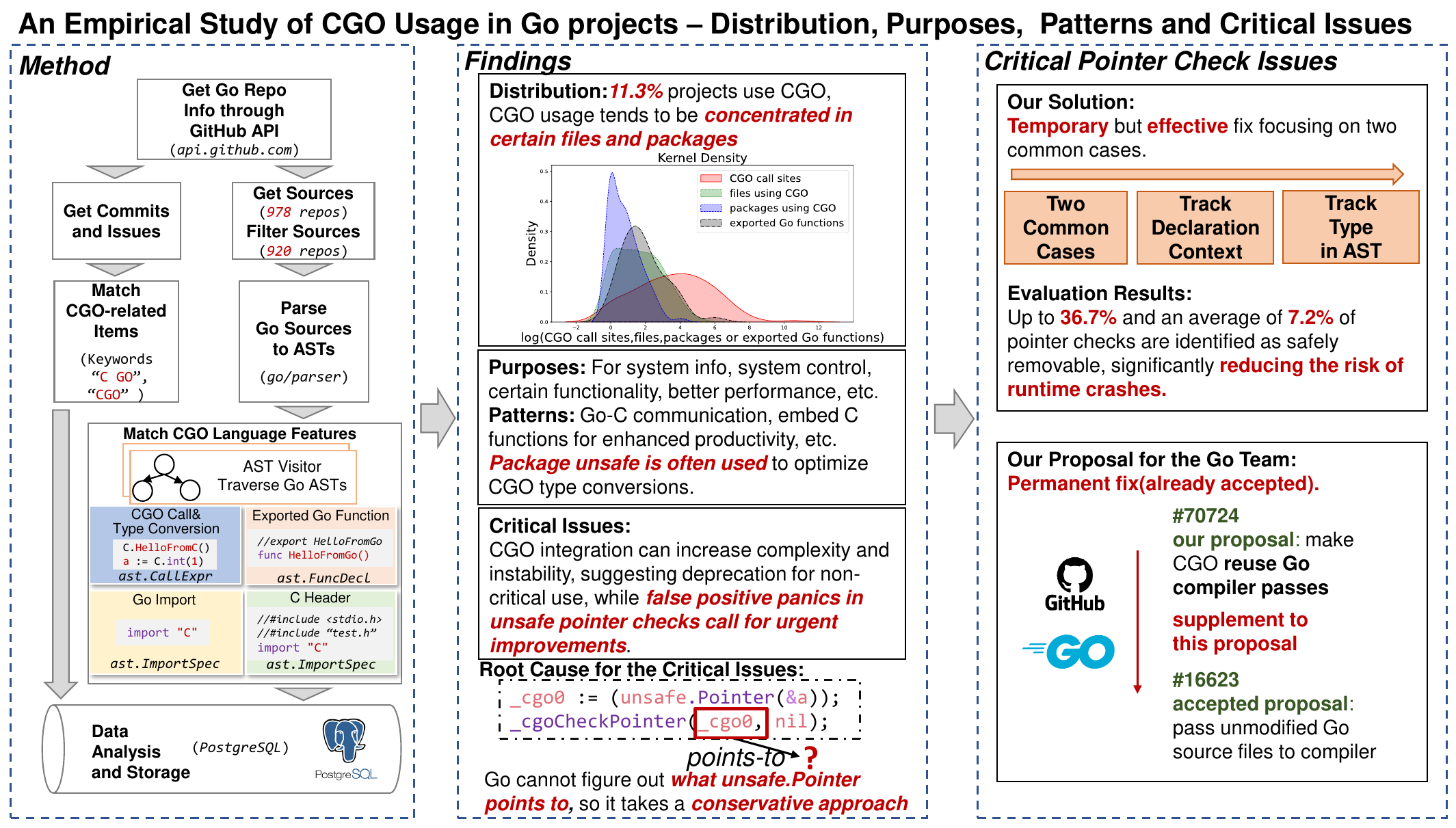}
\end{graphicalabstract}

% 3 - 5 items
% limited in 85 characters per item, including whitespaces.
\begin{highlights}
\item Design \mysys to analyze the usage of CGO, the Foreign Function Interface in Go.
\item Identify CGO usage distribution among projects, along with 4 purposes and 15 patterns.
\item Identify 15 types of CGO issues, including one critical type causing unexpected crashes.
\item Propose approaches to mitigate/resolve the critical issue.
\item Guide developers on reliable CGO use and advise the Go team to make CGO more robust.
\end{highlights}


%% file: intro.tex
\section{Introduction}
\label{sec:intro}
% 开门见山，多语言软件 -> core is 跨语言接口 -> CGO -> ...
Modern software development increasingly embraces multilingual programming, allowing developers to leverage the strengths of multiple programming languages within a single project \remv{(Yang et al., 2024; Mayer et al.,
 2017; Li et al., 2023a; Kochhar et al., 2016).}
%\remv{\citep{yang2024multi, mayer2017xll, li2023multilingual,kochhar2016large}.}
%\citep{yang2024multi, mayer2017xll, li2023multilingual,kochhar2016large}.
%The advantages of  multilingual development are clear, as it enables developers to leverage the unique strengths of various languages, thereby enhancing productivity and flexibility~\cite{mayer2017xll}.
\cjb{~\citep{yang2024multi, mayer2017xll, li2024multilingual,kochhar2016large}.}
%A  multilingual software architecture combines the strengths of the interacting languages, leveraging the advantages of high-level languages, such as ease of use, rich libraries, and rapid development capabilities, while also taking advantage of low-level languages like C for performance-critical tasks, system-level operations, and efficient resource management. 
At the core of multilingual development lies the Foreign Function Interface (FFI), which enables seamless integration and mutual calls between languages. Many traditional programming languages provide FFIs to support cross-language interoperability, such as the Python/C API in Python~\citep{pythoncintro}, JNI in Java~\citep{jniintro} and FFI-NAPI in JavaScript~\citep{ffinapi}. These FFIs greatly boost development efficiency by allowing code reuse, expanding application scenarios, and supporting the creation of more efficient, powerful, and scalable software systems.

%https://en.wikipedia.org/wiki/Foreign_function_interface
%Such interfaces allow the reuse of existing, well-tested, and optimized libraries written in other languages, such as the encryption library, OpenSSL~\cite{openssl} and the compression library, zlib~\cite{zlib} thereby avoiding redundant development work and achieving better time and space efficiency. Popular traditional multilingual projects on GitHub, such as TensorFlow~\cite{abadi2016tensorflow} and PyTorch~\cite{paszke2019pytorch} (Python/C++ for performance), Electron~\cite{electron} (JavaScript/C for backend capabilities) and OpenJDK~\cite{openjdk} (Java/C++ for performance)%go-ethereum~\cite{goethereum} (Go/C for critical operations), and boringcrypto~\cite{boringcrypto} (Go/C using Google's BoringSSL~\cite{boringssl} for cryptography), demonstrate how modern software effectively combines high-level languages for development ease with low-level code.
% 直接写和本文相关的背景，大背景太长了，小背景可以多写。

% 对Go的这些基础介绍也可以简短一些
Go programming language (Golang), which has seen significant growth~\citep{jetbrains2021Go} and reached 7th place in the TOBIE index~\citep{tiobeindex} in December 2024, offers a unique FFI called CGO.
%The Go programming language, designed by Google in 2009, is an emerging programming language. Go has become increasingly popular in recent years, reaching 7th place in the TOBIE index in December 2024, marking its highest position in history~\cite{tiobeindex}.
%\remv{The Go programming language, designed by Google in 2009, is an emerging programming language. It is a statically-typed language and offers a number of language-level features that improve programming efficiency, such as concurrency support at coroutine granularity, garbage collection, simple syntax, etc~\cite{goatgoogle}.}
%\remv{. So Go has become increasingly popular in recent years, In 2015, Go ranked 122nd, however, it quickly rose in popularity, winning the "Programming Language of the Year" award in 2016 and reaching 7th place in December 2024, marking its highest position in history}
%\remv{. Like the aforementioned traditional languages, Go} 
%Go also provides its own unique FFI named CGO~\cite{cgointro}, 
CGO allows direct interaction between Go and C, combining Go's simplicity and efficiency with C's power and maturity. This makes CGO invaluable in areas like system-level programming and performance-critical tasks. With widespread applications of Golang~\citep{jetbrains2021Go}, popular Go projects like Docker~\citep{docker} and Go-Ethereum~\citep{goethereum}, which leverage CGO, are now widely used around the world. 

While FFIs like CGO offer significant advantages in software development, they also introduce challenges. Research\remv{(Li et al., 2023b, 2022; Grichi
 et al., 2020a; Mayer et al., 2017)}\cjb{~\citep{li2023understanding, li2022vulnerable, grichi2020impactjni, mayer2017xll}} shows that differences in language features between programming languages---such as memory management, type systems, exception handling, and concurrency mechanisms---can make cross-language programming more complex, increasing risks in the interface layer. These risks include performance overhead, difficult debugging, and potential security vulnerabilities, etc.~\citep{li2023understanding}. 
%An analysis by ~\cite{li2022vulnerable} of numerous open-source projects and their patch histories found that the greater the number and complexity of languages and interoperability mechanisms, the higher the likelihood of risks in the software system. Current research has also shown that the more interdependencies between languages in a software system, the higher the risk of vulnerabilities, approximately 2–3 times greater than in single-language systems~\cite{grichi2020impact, mayer2017xll}. 

To mitigate these risks, understanding the distribution, patterns, purposes, and critical issues of cross-language is essential for building efficient and secure multilingual systems. Identifying common use cases and pain points can help developers propose better solutions. While much research has focused on traditional Python/C~\citep{hu2020python, hu2023empirical} and JNI~\citep{grichi2020impactjni, hwang2024empirical}, there has been limited study on CGO.
%For traditional programming languages, existing research has conducted in-depth studies on their cross-language interfaces, uncovering relevant usage patterns and associated pain points.
%Therefore, some current research focuses on the study of cross-language interfaces among real-world projects. 
% For example, ~\cite{hu2020python, hu2023empirical} focus on usage statistics and bug patterns in the Python/C API, while ~\cite{grichi2020impactjni, hwang2024empirical} investigate interlanguage dependencies in JNI and the JVM's behavior during erroneous JNI interoperations. 
% Go's unique languauge features:
% It is a statically-typed language and offers a number of language-level features that improve programming efficiency, such as concurrency support at coroutine granularity, garbage collection, simple syntax, etc
%Unfortunately, to the best of our knowledge, most existing research has focused on well-established interfaces like Python/C and Java/C, with no work dedicated to analyzing CGO in Go, a relatively young language. 

Previous studies offer valuable insights, but their findings to other FFIs are not %directly 
applicable to CGO due to Go's unique design philosophy and runtime characteristics. 
Unlike Python or Java, Go compiles programs ahead of time and has simpler language features. Go is a statically-compiled language that avoids using a virtual machine, features a simple syntax, and includes a runtime with built-in concurrency support and other advanced capabilities, setting it apart from Java and Python. 
Go's complex and rich runtime system and CGO's seamless integration with Go’s tooling and runtime strengthen the language's distinctiveness. 

CGO's close coupling with Go allows direct interaction with C code through a pseudo-package (named \kn{C}), simplifying development but introducing unique challenges. For example, Go's garbage collector manages memory automatically, while C requires manual memory management. This discrepancy can lead to memory issues such as dangling references and memory leaks, since Go may not recognize memory allocated and referenced by C. 

To mitigate such memory issues, Go imposes certain restrictions on the use of CGO, such as not allowing Go code to store Go pointers in C memory~\citep{cgopointerpass}. %These restrictions make development more cumbersome and error-prone due to the complexity of type and data conversions and the need to track reference relationships. 
These restrictions require Go to perform checks during compilation or runtime, which may lead to false positives. For instance, Go's issue\#63460~\citep{issue63460} mentions that the current CGO checks may falsely report issues when encountering a struct containing a Go pointer, causing a normal program to crash. Additionally, these complexities make development more cumbersome and error-prone due to the intricacies of type and data conversions and the need to track reference relationships. For instance, in issue\#200~\citep{issue200} of repository \href{https://github.com/go-gl/glfw/issues/200}{\kn{glfw}}, the developer's lack of understanding regarding CGO restrictions and patterns led to incorrect usage, resulting in a runtime panic.
% 这些限制既使得Go Team需要在编译或运行时检查这些限制，这可能导致误报或者漏报，比如Go's issue\#63460提到当前CGO在遇到a struct containing a go pointer时可能导致误报，使得正常程序crash，也使得开发者development more cumbersome and error-prone due to the complexity of type and data conversions and the need to track reference relationships，比如in issue\#200~\cite{issue200} of repository \href{https://github.com/go-gl/glfw/issues/200}{\kn{glfw}}, the developer's lack of understanding regarding CGO restrictions and patterns led to incorrect usage, resulting in a runtime panic.
%Currently, it's still unknown which patterns CGO have.%Additionally, CGO’s reliance on a C compiler creates complexities in cross-compilation and portability, further distinguishing it from other cross-language interfaces. 

Despite CGO's unique characteristics, which may introduce additional challenges and issues, currently, there is a lack of comprehensive study on CGO usage. This gap leaves both Go team and developers ill-equipped to handle its complexities, potentially leading to the runtime errors mentioned above, as well as other avoidable pitfalls such as inefficient code and longer debugging times.%This lack of comprehensive study on CGO usage risks leaving developers ill-equipped to handle its complexities. 
%Without a clear understanding of how CGO is used in practice, developers may encounter avoidable pitfalls, leading to inefficient code, increased debugging time, and potential runtime errors. 
The lack of targeted research also hinders the development of tools and best practices for CGO, limiting its adoption and effectiveness. 

% dby：“to address this research gap”， 我觉得“research gap”不应该作为我们做CGO的一个初始动机。比如上边那一段说了两部分，一部分是CGO实证分析没人做过；另一部分是CGO实证分析的价值和意义。主要我觉得一个研究没人做过并不意味着它有价值，更应该作为研究动机的是它本身的价值而不是没人做过，就是它能给研究或者软件开发带来什么？可以扩写一下“to address this research gap and ...(体现研究价值)”
% 可以写bug的危害等等来表示研究意义。gap不能代表意义。写一些具体例子。
% 
% 我这有一段话不知道能不能用来辅助表示研究价值和研究意义：由于Go语言的内存通过垃圾扫描器管理，在通过CGO与C语言交互时，可能会由于无法识别来自C程序的内存分配和引用而导致悬空引用、内存泄漏等内存安全问题。因此在进行跨语言交互时往往需要繁杂的类型与数据转换以及引用关系的维护和追踪。这不仅使得内存交互效率低，而且还会使得代码变得繁琐且易错。

To address the research gap and provide developers with a comprehensive understanding of CGO, we conduct an empirical study of CGO usage in popular open-source Go projects on GitHub. To the best of our knowledge, this is the first study to reveal the distribution, purposes, patterns, and critical issues associated with CGO, while also proposing solutions to address these identified critical issues. Our findings provide valuable insights for Go developers and the Go team, helping to improve CGO usage and its integration in Go projects. %how and why CGO is used within the Go open-source community, as well as to uncover the potential risks inherent in its adoption.

We developed a tool named \mysys to identify and quantify CGO features in Go projects, offering detailed insights into CGO usage. By leveraging \mysys, we gain a thorough understanding of CGO usage in open-source Go projects. We examine the purposes behind CGO adoption, analyze common usage patterns, and identify potential critical risks it may pose to Go projects, while also proposing solutions to mitigate these risks. Our study is guided by four research questions (RQs): the first three RQs focus on the distribution, purposes, patterns, and common issues, while RQ4 investigates the root causes and solutions for the critical issues identified. The findings from these RQs can help developers write more efficient, safe code and provide valuable insights to the Go team for enhancing the safety of CGO and improving the toolchain robustness.

In summary, we have made the following contributions:

    1) \textbf{Development of \mysys.} We developed \mysys, a tool for identifying and quantifying CGO-related language features in Go projects. It retrieves the top 1,000 most-starred Go repositories via the GitHub API, collects CGO-related issues and commits, and performs static analysis on their ASTs to extract relevant features such as function calls. The results are stored in a centralized database for further analysis.  
    
    2) \textbf{Empirical Study of CGO Usage.} We conducted an empirical study of CGO usage in the retrieved open-source Go projects to uncover its distribution, purposes, patterns, and critical issues. Our study reveals that 11.3\% of projects use CGO, typically concentrated in a limited number of files and packages. We also identify 3 purposes (\S \ref{sec:empirical:rq1}), 15 usage patterns (\autoref{tab:patterns}), and 5 common issue categories (\S \ref{sec:bug}). Notably, false positive runtime crashes caused by CGO's unnecessary pointer checks emerge as a critical issue. 
    
    3) \textbf{Study on Critical CGO Pointer-Check Related Issues.} We found that such critical pointer-check-related crashes stem from Go's inability to determine certain pointer targets during CGO calls (\S \ref{sec:ptrchk:issue}). A permanent fix requires modifying the Go toolchain, a major effort needing Go team's involvement. To address this, we propose a two-phase solution:
    %To fundamentally address this issue, it is necessary to accurately determine the type of the object a pointer points to during CGO calls. %This requires modifying the current Go toolchain %to enable CGO to reuse compiler passes, such as type inference and pointer analysis. However, this is a complex task that we cannot resolve independently and requires support from the Go Team. 
    %This requires modifications to the existing Go toolchain, which is a complex task beyond what we can address independently and would require support from the Go Team.
    %Therefore, in response to this problem, we have taken the following two actions(\S \ref{sec:ptrchk:allsol}):
    \begin{itemize}
        \item%[3.1)] 
        \textbf{Temporary but effective fix:} We designed a lightweight method that tracks pointer types in two common cases. Evaluation results show that up to 36.7\% and an average of 7.2\% of pointer checks are identified as safely removable, significantly reducing the risk of crashes.
        %our simple approach can eliminate up to 36.7\% and an average of 7.2\% of pointer checks, effectively reducing the risk of false positive runtime crashes caused by redundant pointer checks.
        \item%[3.2)] 
        \textbf{Proposal for permanent fix:} We submitted a proposal~\citep{proposal70274} to the Go team, recommending that CGO reuses compiler passes (such as type inference and pointer analysis) to ultimately resolve the issue. Our proposal is currently grouped with an ongoing and previously accepted proposal~\citep{proposal16623}, serving as a supplement by adding additional considerations. % to address this specific problem. 
        Both proposals are expected to make further progress in the future.
        %Our proposal now serves as a supplement to an ongoing and previously accepted proposal. While the two proposals are similar, the previous accepted proposal does not address the issue of false positive crashes caused by CGO pointer checks.
       \end{itemize}
    
    4) \textbf{Implications and Recommendations.} Based on our findings, we provide implications and recommendations for both Go developers and the Go team (\S \ref{sec:proposal}). For instance, developers should carefully assess CGO trade-offs, follow recommended coding styles, and leverage common usage patterns to improve efficiency and mitigate risks. %developers are encouraged to carefully weigh the trade-offs of using CGO and to adhere to the coding styles we provide. They can also draw inspiration from the widely adopted CGO purposes and usage patterns (e.g., using productivity-enhancing wrappers or combining the \codet{unsafe} package to accelerate direct type conversions between Go and C), which can help mitigate risks and improve development efficiency. %For instance, developers are encouraged to carefully weigh the trade-offs involved in using CGO and to adhere to our given coding styles, which can help reduce risks and improve development efficiency.
    The Go team is advised to enhance the pointer-check strategy, addressing current issues to improve the reliability and safety of CGO-based software. These insights offer valuable guidance for developers integrating CGO and suggest enhancements for the Go toolchain.
    %These recommendations aim to create a more robust development experience, optimize runtime behavior, and support the efficient and reliable use of CGO in Go projects. 

This paper is organized as follows: In \S \ref{sec:related}, we provide the background for our paper and introduce related work. In \S \ref{sec:methodology}, we introduce the four research questions (RQs) that this paper aims to answer and describe how we leverage our \mysys tool to conduct both quantitative and qualitative analyses. Then, we first focus on the study of CGO usage to address the first 3 RQs in \S \ref{sec:empirical}. In \S \ref{sec:ptrchk}, to address RQ4, we analyze the root causes of the critical issues identified in RQ3 and propose both a temporary solution and a proposal for a permanent fix. We also evaluate our approach and provide further discussion. Some implications for Go developers and Go team are given in \S \ref{sec:proposal} according to our findings. Finally, we summarize potential threats to validity in \S \ref{sec:threats} and offer conclusions in \S \ref{sec:conclusion}.

%% file: related.tex
\section{Background and Related Work}
\label{sec:related}
%\dbynote{Focus more on introducing the significance of the research, rather than overly emphasizing that no one has done this research before.}
In this section, we highlight how CGO's unique characteristics introduce distinct challenges, which may pose potential risks to development (\S \ref{sec:pre}). We also emphasize the gap in existing research, as most studies fail to address CGO-specific issues (\S \ref{sec:relatedwork}). 
%Specifically, in \S \ref{sec:related:multilang-dev}, we will introduce current studies from the broader perspective of multilingual software development to highlight the risks and complexities inherent in multilingual software, which require further investigation. In \S \ref{sec:related:multilang-interface}, we will narrow the focus to FFIs, which are key to implementing multilingual software, and discuss current research on specific FFIs, such as Python/C API and JNI, while pointing out the lack of research on CGO. In \S \ref{sec:related:go}, we will briefly outline the current work on Go, which primarily focuses on concurrency and type safety, but not on the CGO interface. 
%Finally, in \S \ref{sec:pre}, we briefly introduce the unique usage and processing flow of CGO as background information, along with the unique challenges it brings. We also explain why previous research on cross-language interfaces like the Python/C API and JNI cannot be directly applied to CGO, setting the stage for our study on CGO.

\input{pre}

\subsection{Related Work}
\label{sec:relatedwork}
\subsubsection{Studies on Multilingual Software Development}
\label{sec:related:multilang-dev}
\citet{li2022vulnerable} conduct a study of multilingual code on GitHub, revealing statistically significant associations between vulnerability proneness and language selection as well as interfacing mechanisms, and provide recommendations to mitigate multilingual security risks. \remv{Grichi et al. (2020a)}\cjb{\citet{grichi2020impactjni}} analyze dependencies in multilingual systems, revealing that inter-language dependencies significantly increase the risk of bugs and vulnerabilities compared to intra-language dependencies. \citet{mayer2017xll} present an empirical survey of 139 professional developers, revealing that while multilingual development enhances motivation and requirement translation, it also poses significant challenges in understandability, changeability, and cross-language linking, etc. \citet{li2024multilingual} conduct a characterization of language use and selection in multilingual software, revealing trends, functional associations, and evolutionary shifts in language profiles. \citet{yang2023demystifying, yang2024multi} analyze Stack Overflow posts to uncover key challenges in multilingual software development, including interoperability complexities, error handling, and gaps in language-specific technical knowledge, offering actionable insights for researchers and developers. \citet{li2023understanding} present a study of several bugs and multilingual projects, revealing that multilingual bugs are more complex, take longer to resolve, and have higher reopen rates than single-language bugs. \citet{grichi2024change} investigate Change Impact Analysis (CIA) in multilingual systems, revealing a strong industry demand for CIA tools

\subsubsection{Studies on FFIs}
\label{sec:related:multilang-interface}
As the core of multilingual development, FFIs play a critical role in enabling the integration of different programming ecosystems. Study on FFIs is crucial because they directly impact software interoperability, performance, and maintainability. Understanding their design, usage patterns, and limitations helps developers make informed decisions about when and how to leverage them effectively. \citet{hu2020python, hu2023empirical} investigate the Python/C API, highlighting its evolution, usage statistics, and bug patterns with empirical examples from real projects, and propose a systematic bug taxonomy to guide the development of automated, high-precision bug-finding tools. \citet{monat2021multilanguage} introduce a static analyzer that fully automates the analysis of Python programs calling C extensions, detecting runtime errors in Python, C, and at the interface. \citet{grichi2020impactjni} conduct an empirical study on ten Java Native Interface (JNI) systems to identify the interlanguage dependencies, analyze their prevalence in multilingual systems, and their impact on software quality and security. \citet{hwang2024empirical} focus on the challenges of developing correct JNI programs and presents a semi-automatic tool designed to generate JNI test programs. \citet{ding2023cgorewritter} propose \kn{CGORewriter}, a tool aimed at simplifying the integration of C libraries into Go via CGO. This is currently the only paper related to CGO, as it focuses on tool development rather than an in-depth analysis of the CGO interface itself.

%Despite the extensive research on Python/C API, JNI, or other multilingual systems, there is a lack of studies focusing specifically on CGO. Unlike Python/C or JNI, CGO allows Go programs to directly call C code, and its integration mechanism is unique, for example, it requires both Go's runtime and C's memory management to coexist, often introducing specific challenges in memory handling, concurrency, and error propagation, etc. Previous work on Python/C or JNI cannot be directly applied to CGO due to these distinct differences in design. 

%The absence of focused studies on CGO presents a significant gap in understanding its potential pitfalls, particularly in terms of performance optimization, error handling, and security vulnerabilities in multilingual Go applications. Without research dedicated to CGO, developers may face increased difficulty in creating efficient, secure, and maintainable Go/C applications, leading to potential misuse of the interface, overlooked bugs, and security risks that could impact the reliability of multilingual systems.

% {\bf Empirical studies on Go.}
\subsubsection{Studies on Go}
\label{sec:related:go}
The current study on Go mainly focuses on concurrency~\citep{ng2016goDeadlock, lange2017go, lange2018go, dilley2019Go, tu2019go, gabet2020static, liu2021gcatchGfix, liu2022goes, saioc2024unveiling} and type safety~\citep{lauinger2020unsafeGo, costa2021breaking, wickert2023ungoml}, with little attention given to CGO, except for \kn{CGORewriter}~\citep{ding2023cgorewritter} mentioned before. For example, \citet{dilley2019Go} analyze 865 Go projects from GitHub to better understand how message passing concurrency is used in practice.
\citet{tu2019go} study Go concurrency bugs and divide them into blocking bugs (\eg deadlock) and non-blocking bugs (\eg data race).
Both \citet{lauinger2020unsafeGo} and \citet{costa2021breaking} study the usage of \kn{unsafe} package in popular 500 or 2,438 Go projects, respectively. \citet{dilley2021go} propose a static checker for concurrent safety, while \citet{lange2017go, lange2018go} develop a framework to detect communication errors and deadlocks in Go programs. \citet{ng2016goDeadlock} use state machines to detect deadlocks. Up to now, as far as we know, there is no study concentrated on CGO.

%% file: pre.tex
\subsection{Background - CGO}
\label{sec:pre}

CGO is an %\remv{cross-language calling mechanism between Go and C} 
FFI provided by Go, enabling the interoperability between Go and C%\remv{to call each other}
~\citep{cgointro}. %the creation of Go packages that call C code.

\subsubsection{Unique Usage}

%Go natively provides a cross-language call mechanism between Go and C named CGO. 
%\del{In Go, as shown in \autoref{code:cgo:go2c}, we enable the CGO mechanism simply by adding \codet{import "C"} statement in the Go code~(Line 6 ), and the C code can be embedded in Go code as a preamble (Lines 2 to 5). Then the C function can be called in Go from a virtual package ``\codet{C}''~(Line 10).}
In Go, as depicted in \autoref{code:cgo:go2c}, CGO can be enabled by simply adding the \codet{import "C"} statement~(Line 6). C code can be embedded in Go code within comments as a preamble~(Lines 2 to 5), allowing C functions to be called in Go directly through a virtual package named \codet{"C"}~(Line 10).
% 这样的注释有个术语叫 preamble
%\autoref{code:cgo:go2c} is an example of calling C function \kn{print\_ptr} from Go. 

\begin{figure}[htb]

\centering
\begin{minipage}{.45\textwidth}
\begin{lstlisting}[style=Go]
package main
/*#include <stdio.h>
void print_ptr(void *p) {
	printf("%p\n", p);
}*/
import "C"
\end{lstlisting}
\end{minipage}%
\hfill
\begin{minipage}{.45\textwidth}
\begin{lstlisting}[style=Go, firstnumber=last]
import "unsafe"
func main() {
	var a C.int = 1
	C.print_ptr((unsafe.Pointer(&a)))
}
\end{lstlisting}
\end{minipage}

\caption{PrintGoPtr.go}
\label{code:cgo:go2c}

\end{figure}

% \begin{center}
% \begin{minipage}[c]{0.469\textwidth}
% \lstinputlisting[
%     style       =   Go,
%     caption     =   {\bf PrintGoPtr.go},
%     label       =   {code:cgo:go2c}
% ]{sections/src/cgo/HelloFromC.go}
% %\vspace{-0.3cm}
% \end{minipage}
% \end{center}

Go functions can also be invoked from C by exporting Go functions to libraries. To export a Go function, simply add a comment with the \codet{export <funcname>} directive before the function (Line 3 in \autoref{code:cgo:c2go_go}). The Go compiler will export the function to the library, making it callable from C code (Line 3 in \autoref{code:cgo:c2go_c}).
% not statement, 一般叫做 directive
%\autoref{code:cgo:c2go_go} and \autoref{code:cgo:c2go_c} are examples of calling Go functions from C. 

\begin{center}
\begin{minipage}[c]{0.4\textwidth}
\begin{lstlisting}[style=Go]
package main
import "C"
//export HelloFromGo
func HelloFromGo(){
	println("From Go") }
\end{lstlisting}
\captionof{figure}{\bf HelloFromGo.go}
\label{code:cgo:c2go_go}
\end{minipage}
\hfill
\begin{minipage}[c]{0.4\textwidth}
\begin{lstlisting}[style=Go]
#include "HelloFromGo.h"
int main(){
    HelloFromGo();
    return 0;
}
\end{lstlisting}
\captionof{figure}{\bf HelloFromGo.c}
\label{code:cgo:c2go_c}
\end{minipage}
\end{center}

CGO integrates seamlessly with Go’s build system, in contrast to other FFIs like the Python/C API and JNI, which require %This design contrasts with the Python/C API, JNI, and other existing FFIs, where 
explicit setup and interface definitions %are required
to bridge the language boundary. For example, the Python/C API necessitates defining Python-compatible modules, functions, and data structures in C, along with careful reference counting to manage memory. %Similarly, JNI requires Java methods to be declared as native, and corresponding implementations in C or C++ must adhere to the JNI conventions, including signature matching and JVM interaction. 
In contrast, CGO %integrates seamlessly with Go’s build system, requiring 
requires minimal boilerplate and abstracts much of the complexity of marshaling and unmarshaling data across language boundaries, whereas the Python/C API and JNI demand more manual handling. This simplicity and tight integration make CGO distinct. Moreover, Go's primary application domains, such as websites, utilities, and IT infrastructure~\citep{jetbrains2021Go} may differ from other languages such as Python and Java. 

These fundamental differences in usage mean that previous studies on usage statistics, patterns, or other factors for the Python/C API, JNI, or other FFIs may not directly apply to CGO. The streamlined integration and distinct usage of CGO result in unique usage characteristics that require independent study. 

% Therefore, to gain a better understanding of CGO usage, we need to conduct a comprehensive study of how CGO is used in projects, including its distribution, usage purposes, and other relevant factors

%As a result, the purposes and motivations behind cross-language interfacing in CGO are distinct, further limiting the relevance of prior research to this context. 

% //this header is generated by cmd/cgo automatically

%% 我觉得这段不需要，而且会显得像技术笔记
% To run this example, we can use commands below to wrap Go function into a static library and then ld the library with the C code.

% \lstinputlisting[
%     style       =   Shell
% ]{sections/src/cgo/command.sh}
% \begin{lstlisting}[language=Bash]
% go build -buildmode=c-archive -o HelloFromGo.a HelloFromGo.go
% gcc HelloFromGo.c HelloFromGo.a -lpthread 
% ./a.out
% \end{lstlisting}

\subsubsection{Unique Restriction}
\label{sec:bg:restriction}
%Go is a garbage-collected language, and the garbage collector needs to know the location of every pointer to Go memory. Because of this, there are restrictions on passing pointers between Go and C~\cite{cgopointerpass}. 
CGO enables mutual function calls between Go and C, allowing Go's runtime components, such as the garbage collector, memory allocator, and goroutine scheduler, to coexist with C during execution. To ensure the proper functioning of Go's runtime components, such as the garbage collector, Go imposes certain restrictions that developers must follow when passing pointers between Go and C using CGO~\citep{cgopointerpass}.
%certain restrictions apply when passing pointers between Go and C in CGO~\cite{cgopointerpass}. 
For example, Go code is not allowed to pass any pointers that point to Go memory containing Go pointer, nor to store Go pointers in C memory.
% C code may store Go pointers in C memory but must stop storing the Go pointer when the C function returns.
These restrictions can be checked dynamically at runtime by the function \codet{\_cgoCheckPointer} inserted by the Go compilation toolchain.
% 栈扩展/栈移动的影响

Moreover, due to the differing runtime management systems of Go and C, developers must comply with certain implicit restrictions to ensure the proper functioning of programs. For example, since Go manages memory using a garbage collector while C relies on manual management, interactions between Go and C via CGO can result in memory issues like dangling pointers and memory leaks, as the Go runtime cannot recognize memory allocations and references in C. %Consequently, cross-language interactions often require additional data structures and processing to prevent such issues. Developers need to be aware of these implicit constraints to ensure the reliability of their code.

Such restrictions require developers to redesign data representations and adapt third-party libraries, involving complex type and data conversions, as well as the maintenance and tracking of reference relationships when using CGO. This makes the code more cumbersome and error-prone, creating challenges that are not present in well-established interfaces like the Python/C API or JNI. 

Unfortunately, there is currently no dedicated study on CGO, leaving developers unaware of its restrictions. This can have a negative impact on development, both in terms of efficiency and reliability. For instance, in issue\#200~\citep{issue200} of repository \href{https://github.com/go-gl/glfw/issues/200}{\kn{glfw}}, the developer's lack of understanding regarding CGO restrictions and patterns led to incorrect usage of CGO, causing the program to panic during runtime. % 该issue花费了xxx

Therefore, To make developers aware of these explicit and implicit restrictions, and to identify the best development practices that can enhance both efficiency and reliability, a comprehensive study of the current CGO usage patterns and issues is necessary.

\subsubsection{Unique Process Flow}
\label{sec:intro-compilaton}
For CGO programs, the Go compilation toolchain performs additional preprocessing steps. It rewrites the original CGO calls to generate multiple intermediate files, which are then compiled along with the rest of the program during the compilation stage. During this process, the default C compiler (usually \kn{GCC}) is invoked to handle the compilation of the C code. This is in contrast to Python/C API and JNI, where the source code typically doesn't require such preprocessing and instead focuses on runtime bindings. CGO directly integrates Go and C, managing both the Go and C parts together.

%During rewriting, Go generates a corresponding Go function signature for each C function. Then the Go compiler checks each CGO call. If it identifies that the argument's type might contain a pointer to a Go pointer, it inserts a function called \codet{\_cgoCheckPointer} to verify whether the argument indeed has such a pointer at runtime. \myins{If such a pointer is found, a runtime panic is triggered.} %\del{Currently there are some issues in the Go community about false positive runtime panic caused by \codet{\_cgoCheckPointer}, for example, issue\#14210~\cite{issue14210}.}

Additionally, during runtime, CGO handles challenges like ABI conversion, translating data types and calling conventions between the two languages, and stack switching, seamlessly transitioning between Go stack~\citep{gostack} and C stack. It also synchronizes Go's runtime with C calls to prevent conflicts, such as pausing garbage collection when necessary. Unlike Python/C API, which requires manual reference counting and memory management, CGO operates more automatically by integrating closely with the Go runtime. Such unique automatic runtime management ensures that C functions do not disrupt Go’s runtime system.

However, there is currently a lack of dedicated studies on CGO, leaving developers without a clear understanding of the potential issues this unique process flow might introduce. Without such understanding, developers may unknowingly implement inefficient or error-prone solutions, which could negatively impact development efficiency and reduce the reliability of software.

%% file: methodology.tex
\section{Methodology}
\label{sec:methodology}

In this section, we propose four research questions (RQs) in \S \ref{sec:method:rq} to explore CGO's usage statistics, purposes, patterns, problems, and critical issues. To automate data (code) collection and analysis needed to answer these RQs, we design a tool named \mysys (\S \ref{sec:mysys}), which automates the extraction of CGO-related data from Go projects on GitHub. Additionally, addressing these RQs requires qualitative analysis of certain data, such as issues identified by \mysys and their code patterns. In \S \ref{sec:qualitative}, we outline our methodology for conducting scientific qualitative analysis.

\subsection{Research Questions}
\label{sec:method:rq}

\textbf{RQ1: How often is CGO used in real-world Go projects, and for what purposes?} This RQ aims to explore the distribution of CGO usage and identify the main reasons developers adopt it. By analyzing the frequency and context of CGO usage, we seek to understand its prevalence in practice and the key motivations behind its adoption. The insights gained from this RQ provide a foundational understanding of CGO's role in modern Go software development.

\textbf{RQ2: What are the most frequently used CGO language patterns, why are certain patterns preferred, and how are they applied?} This RQ seeks to identify the CGO language patterns most favored by Go developers and analyze the reasons behind their popularity. We aim to understand how these patterns are applied in Go projects, focusing on their impact on safety, performance, convenience, and flexibility. By examining these factors, we aim to provide insights into the selection and implications of CGO patterns, helping developers recognize best practices and avoid common pitfalls in CGO usage.

\textbf{RQ3: What are the common problems when using CGO, and which of these are critical issues?}
This RQ aims to identify frequent problems encountered with CGO, focusing on critical issues that could lead to safety risks, performance degradation, or stability concerns in software. By analyzing these challenges, we seek to highlight the most significant issues, helping developers understand the potential risks associated with CGO and how to mitigate them effectively.

\textbf{RQ4: What causes these critical issues, and how can they be mitigated?} This RQ aims to uncover the root causes of the critical issues identified in RQ3 
and explore potential methods to mitigate them. We propose practical solutions to help developers overcome these challenges, and submit proposals to the Go team to address these critical issues from a language design perspective.

\subsection{Design of \mysys{}}
\label{sec:mysys}

\begin{figure}[htb]
    \centering
    \includegraphics[width=1.0\linewidth]{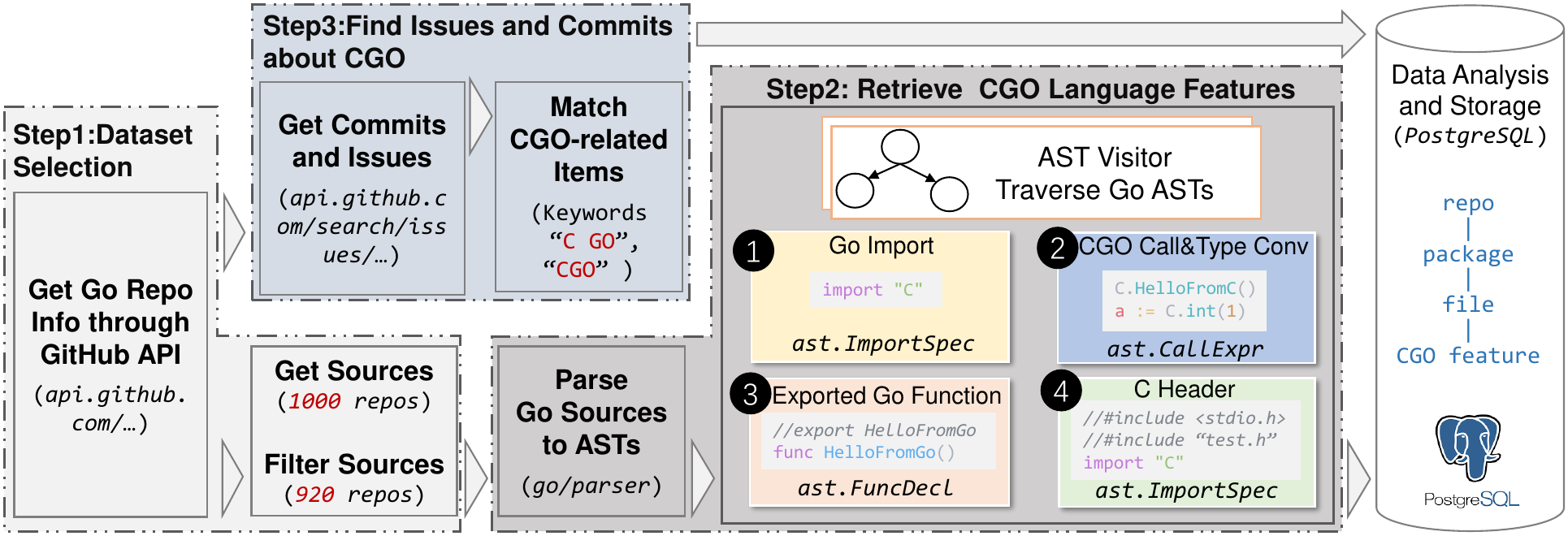}
    \caption{Architecture of \mysys{}}
    \label{fig:cgoanalyzer}
\end{figure}

\mysys{} is an automated tool we designed to identify and quantify CGO language features in open-source Go projects on GitHub. \autoref{fig:cgoanalyzer} shows the overview structure of \mysys. It is mainly divided into three steps.

\subsubsection{Step1: Dataset Selection}
To study CGO usage in real-world Go projects, we collected open-source projects from GitHub. \mysys leverages the GitHub API~\citep{githubapi} to select a dataset from highly-starred Go repositories, with the star count serving as an indicator of popularity. The initial dataset comprised the top 1000 most-starred Go repositories, ranging from 2159 to 90488 stars. \mysys filtered duplicates, yielding 1000 unique open-source repositories.
%\del{To understand how CGO is used in real-world Go projects, we collect open-source projects in GitHub as the data source for our empirical research. \mysys uses the GitHub API to select dataset from GitHub open-source projects. Our dataset contains the top 1000 most starred Go repositories, and the star number is a good reflection of the popularity of a repository. }
%We write a script to 
% 随后，为了保证我们数据的有效性和时效性，我们又对我们的数据集进行了如下的筛选：

To ensure the validity and timeliness of the data, we applied the following filters:
    \normalsize{\textcircled{\scriptsize{1}}}\normalsize \kw{Archived Projects:} We exclude 37 archived projects which are no longer maintained.
    \normalsize{\textcircled{\scriptsize{2}}}\normalsize \kw{Educational Projects:} We manually filter out 24 educational projects.
    \normalsize{\textcircled{\scriptsize{3}}}\normalsize \kw{Outdated Projects:} We screen out 19 projects last updated before January 1, 2020 \cjb{, to ensure alignment with modern Go practices, particularly the adoption of Go Modules (standard since Go 1.13 in 2019). This cutoff was validated by checking the \kn{updated\_at} timestamps of the remaining repositories, all of which were last updated on or after March 23, 2023.}
After performing the filters above, we end up with 920 pieces of repository information. We finally use \codet{git clone} to clone all the 920 repositories.

\subsubsection{Step 2: Retrieve CGO Language Features}
The CGO language features encompass both utilizing C in Go and incorporating Go in C. These features can be extracted from the Abstract Syntax Tree (AST). 
%\del{To identify every CGO language feature in a Go project, \mysys traverses the Go project's AST (abstract syntax tree) to find them and then records them in a database. }
Hence, to pinpoint every CGO language feature within a Go project, \mysys navigates through the project's Abstract Syntax Tree (AST) to locate them, subsequently documenting them in a database. We use the \kn{go/parser}~\citep{goparser} package of Go to build standard Go AST from Go source files and use \kn{PostgreSQL}~\citep{postgresql} as our database. \mysys implements an \codet{ast.Visitor} interface from package \kn{go/ast}~\citep{goast} to walk through the Go AST. 
\mysys identifies four categories of CGO language features, corresponding to \ding{182}-\ding{185} in \autoref{fig:cgoanalyzer}, covering both Go-to-C calls and exported Go functions callable by C.

    \ding{182} \kw{Detection of CGO:} \mysys inspects each \codet{ast.File} node in the AST to determine whether the \codet{Imports} field contains the \codet{import "C"} directive. Projects with files containing this declaration are classified as utilizing CGO, while those without it are categorized as not using CGO.
    
    \ding{183} \kw{Detection of Go-to-C Calls and CGO Type Conversions:} In Go, function calls and type conversions are represented as \codet{ast.CallExpr} nodes in Go AST. \mysys analyzes these nodes to identify Go-to-C calls and CGO type conversions. It checks if the node implements the \codet{ast.SelectorExpr} interface and if the package name in field \codet{Sel} is \codet{C}. When the function name matches a predefined list of C type names, the node is classified as a CGO type conversion; otherwise, the node is identified as a Go-to-C function call. 
    
    \ding{184} \kw{Detection of Exported Go Functions for C-to-Go Calls:} Go functions that are callable from C must be explicitly exported.
    \mysys identifies such functions by inspecting each  \codet{ast.FuncDecl} node for a \codet{//export xxx} directive in its \codet{Doc} field. 
    The presence of this directive indicates that the function is exported for C-to-Go interaction. 
    
    \ding{185} \kw{Detection of Embedded C Code and Compiler Directives:}
    In Go code files utilizing CGO, embedded C code and related directives (\eg  \codet{\#include} for headers and \codet{\#cgo} for C compiler/linker parameters (see \autoref{code:pattern:build})) are included in the \codet{ast.CommentGroup} node of the \codet{import "C"} statement. \mysys analyzes these nodes to extract information about the included headers and linked libraries for further qualitative analysis in RQ1 to RQ4.

In our analysis, we excluded folders named ``\codet{vendor}'' or ``\codet{test}'', as well as packages in Go projects whose names end with ``\codet{\_test}'', because we focus on code of these projects themselves.

After gathering all CGO language features, \mysys organizes and stores the data in our predefined database, using various tables for repositories, files, CGO calls, etc., linked via foreign keys. This setup enables efficient statistical analysis through SQL queries.

\subsubsection{Step 3: Find Issues and Commits about CGO}
\label{sec:mysys:issues}
\mysys utilizes the GitHub API to search for issues and commits related to ``\kn{CGO}'' or ``\kn{C GO}'' (case insensitive) keywords in those repositories that use CGO identified in Step 2. We manually rule out issues within the \kn{golang/go} repository, which is the official Go source code repository and issue tracker, as our primary focus is on how CGO is used in open-source repositories rather than the internal development and evolution of CGO within the Go toolchain. Those matched entries are then manually reviewed to identify valuable ones for further analysis of CGO usage and related issues encountered by developers. Finally we obtain 143 valid issues.

\subsection{Qualitative Analysis}

\label{sec:qualitative}
In RQ1 to RQ4, qualitative analysis is sometimes necessary, requiring annotations on certain entries such as files, modules, and issues. To ensure the reliability of the qualitative analysis, we employed the open card-sort method~\citep{fincher2005making} for annotation. Initially, both annotators independently review and categorize the data, assigning preliminary labels based on observed patterns. When new labels emerged, our authors engaged in discussions to refine and standardize the categories, ensuring consistency across annotations.

To further validate the annotation process, we assess inter-annotator agreement using the Cohen-Kappa metric~\citep{mchugh2012interrater}, which quantifies the level of agreement beyond chance. In cases where discrepancies arise, both annotators conduct follow-up discussions to reconcile differences, refine label definitions, and consolidate the results. This iterative approach helps us maintain annotation accuracy and minimizes subjective bias.

Taking the issue analysis in Step 3 as an example, we label issues based on multiple factors, including the issue title, description, subsequent comments, and any related pull requests (PRs). Initially, both annotators independently review each issue to determine its primary category (e.g., issues related to project compilation and build processes). Additional labels are then assigned based on specific details, such as relevant comments or PRs. For instance, if an issue also involves dependency management, it may be given the \kn{Dependency} label. If an issue spans multiple concerns, we assign multiple labels accordingly. For ambiguous cases, both annotators engaged in further discussion to refine the labeling criteria and reach a consensus. 

%Through this structured annotation process, we aim to systematically classify issues and extract meaningful insights from the dataset.

%% file: rq.tex
\input{rq1}

\input{rq2}

\input{rq3}

%% file: rq1.tex
\section{Study Results on RQ1 to RQ3 - CGO Usage Analysis}
\label{sec:empirical}
We then perform an empirical study about CGO usage in Go projects using \mysys to first 
 answer RQ1$\sim$RQ3.
\subsection{RQ1: Distribution and Purposes of CGO Utilization}
\label{sec:empirical:rq1}

\mysect{Frequency of CGO Usage.}
Using \mysys, we statistically derive the frequency of cross-language calls between C and Go in open-source projects from our dataset. As shown in \autoref{rq1:repo-num}, out of 920 projects analyzed, 104 projects (11.3\%) utilize CGO, and 33 projects (3.2\%) exhibit more than 100 CGO function call sites per project. These CGO functions encompass both C functions called by Go and built-in Go functions tailored for CGO.

\begin{table}[htb]
\centering
\small
\caption{Number of projects that use CGO}
\scalebox{0.9}{
\begin{tabular}{l|l|l} % <-- Alignments: 1st column left, 2nd middle and 3rd right, with vertical lines in between
\hline
      \textbf{Statistical Range} & \textbf{Number} & \textbf{Percentage}\\
    %   $\alpha$ & $\beta$ & $\gamma$ \\
      \hline
      All projects & \mydel{968}\myinsert{920} & 100\%\\ \hline
      Projects using CGO & \mydel{135}\myinsert{104} & \mydel{13.95}\myinsert{11.3}\%\\ \hline
      Projects using CGO more than 100 times & \mydel{53}\myinsert{33} & \mydel{5.48}\myinsert{3.2}\%\\ \hline
\end{tabular}
}
\label{rq1:repo-num}
\end{table}

% 图一中的三张图显示了在使用了CGO的开源库中，使用CGO的数量以及使用CGO的文件、package的数量分布。首先对于图一，我们可以看到多数库使用CGO的数量在10到1000之间，占比为61.69%，而引用CGO次数大于1000的库则比较少，占比为4.54%。其次，对于图二和图三，我们可以看到在大部分的项目中，CGO的使用主要分布在一个包中以及数个文件中。于此同时，我们也对平均每个项目中所含有的代码行数、包数以及文件数量进行了统计，结果如表格二所示。结合图一中的分布情况，我们不难发现，开源项目对CGO的使用通常都集中在项目的某个或某几个包中，并且含有CGO的文件\包通常Go代码量都比较小（因为使用CGO的包和文件占比分别达到了4.25\%和3.68\%，而使用CGO的代码行数占比仅0.32\%）。这不难理解，因为使用CGO的文件中，多数的核心代码是使用C来实现的，这使得CGO相关的Go代码占比较少。

\input{CGO-RQ1}

\mysect{Distribution of CGO Usage.} \myins{To analyze the distribution of CGO function call sites, packages, and files per project, we use logarithmic counts (base $e$) and plot a Kernel Density Estimate (KDE) curve (see \autoref{rq1:kde}). The area under the curve is normalized to 1, with the Y-axis representing distribution density. The KDE plot shows that most CGO projects (60.58\%) have between 10 and 1000 CGO function call sites, with a small percentage (5.8\%) exceeding 1000 CGO call sites per project. Additionally, most projects use CGO within a single package and fewer than 10 files. The number of projects exporting Go functions for C typically ranges from 4 to 20.}

\begin{table}[!ht]
\centering\small
\caption{Average number of items and CGO-related items}
\scalebox{0.95}{
\begin{tabular}{l|l|l|l|l|l} % <-- Alignments: 1st column left, 2nd middle and 3rd right, with vertical lines in between
\hline
      \textbf{} & \textbf{Total(Go+C)} & \textbf{CGO} & \textbf{C} & \textbf{$P_{CGO}$} & \textbf{$P_C$}\\
    %   $\alpha$ & $\beta$ & $\gamma$ \\
      \hline
      \kd{Lines of Code} & 40422.81 & 62.84 & 3023.74 & 0.16\% & 7.48\%\\ \hline
      \kd{Packages} & 58.26 & 0.41 & - & 0.71\% & - \\ \hline
      \kd{Files} & 213.53 & 1.31 & 17.40 & 0.62\% & 8.2\%\\ \hline
\end{tabular}
}
\label{rq1:pkg-file}
\end{table}

We also analyze the average number of lines of code (LOC), packages, and files in each project (see \autoref{rq1:pkg-file}). Columns 3 and 4 represent Go code with CGO function calls, and code in C source files, respectively, while Columns 5 ($P_{CGO}$) and 6 ($P_C$) show the percentages compared to Column 2. Combining this with the distribution in \autoref{rq1:kde}, we observe that CGO usage in open-source projects is concentrated in one or a few packages, with relatively small amounts of Go code containing CGO calls. Specifically, CGO code accounts for only 0.16\% of LOC, whereas C code comprises 7.48\% of LOC, highlighting that core CGO functionality is primarily implemented in C, leading to lower LOC ratios for CGO function calls.

% 在这些开源项目中，使用CGO频次最高的项目是gl，它是一个针对 OpenGL 的 Go 语言绑定，它含有39774次CGO使用，其中包含了19次的C对Go的调用和39755次的Go对C的调用。这个项目之所以会有这么多的CGO的引用是因为它为不同版本的OpenGL都单独实现了一个Go语言的绑定，每个版本都会包含数千次的引用。通过深入的人工分析发现，项目gl中大部分对于CGO的使用就是通过一个包装函数将OpenGL中的库函数包装成Go函数，大致用法如代码1所示。可以发现，写这种包装函数的工作是大量、机器且重复的。所以值得关注的是，gl库中的绑定代码是通过一个名为glow的生成器生成出来的。glow通过一系列的配置文件并解析OpenGL XML API registry和EGL XML API registry来自动化的生成针对某个OpenGL版本的Go绑定。这意味着使用自动化的程序来通过CGO生成接口代码是可行的。
\begin{center}
\begin{minipage}[c]{0.7\textwidth}
%\centering
\begin{lstlisting}[style=Go]
func BindFramebuffer(target Enum, fb Framebuffer) {
	C.glBindFramebuffer(target.c(), fb.c()) }
\end{lstlisting}
\captionof{figure}{An example of using CGO in \kn{gl}}
\label{code:rq1:example}
%\label{rq1:example}
\end{minipage}
\end{center}

\mysect{Features of the \cjbr{project}{Project} with the Most CGO function calls}.
The \href{https://github.com/go-gl/gl}{\kn{gl}} project, a Go binding for OpenGL, contains 19 C-to-Go calls and 39755 Go-to-C calls, totaling 39774. The project provides bindings for different versions of OpenGL, with each version containing thousands of references. We find that most CGO usage in project \kn{gl} is dedicated to wrapping the OpenGL library functions into Go functions through a wrapper. As shown in \autoref{code:rq1:example}, writing this wrapper function is a mechanical, repetitive task. This repetitive task is handled by \href{https://github.com/go-gl/glow}{\kn{glow}}, a semi-automatic generator that creates CGO \cjbr{bridge}{bridging} code from OpenGL's XML API registries. This highlights the potential of automating the generation of bridging code between Go and C, simplifying the development workflow.

\mysect{\cjb{Usage of (Semi-)Automated Tools for CGO Bridging Code Generation.}}
\cjb{The use of \href{https://github.com/go-gl/glow}{\kn{glow}} in the \kn{gl} project is not an isolated case. Following the qualitative analysis method in \S \ref{sec:qualitative}, we manually inspect all 104 CGO projects in our dataset to investigate the prevalence of such automation. We identified tool usage based on evidence found in its documentation and code comments, which often attribute the generation of CGO bridge code to a specific tool. Our analysis reveals that 10 of the 104 projects (9.6\%) employ automated tools. We categorize their usage into three types (\autoref{tab:cgo-automation-tools}): }

\cjb{\kw{1) Comprehensive binding generation (Full):} For large and complex C libraries, manually writing CGO wrappers is exceedingly tedious. Some projects, including \kn{gl} and \kn{glfw}, \kn{qt} and \kn{pan-light}, use custom or template-based generators to create bindings. Similarly, \kn{nuklear} uses the \href{https://github.com/xlab/c-for-go}{\kn{c-for-go}} tool based on a configuration file, while \kn{macdriver} uses a multi-step process to generate Go wrappers for Apple's Objective-C frameworks. }

\cjb{\kw{2) Specific task automation (Partial)}: %A more subtle but equally important use of automation is to handle specific tasks in CGO development. 
Some projects automate specific, repetitive, or complex subtasks rather than the entire binding process. For instance, \kn{cockroach} uses a custom tool to generate CGO stub files that handle complex and platform-dependent CGO build directives. Projects like \kn{beats} and \kn{confluent-kafka-go} automate the conversion of C error codes into Go representations, reducing manual effort and minimizing error-handling logic. }

\cjb{\kw{3) Tool Provider for the Community (Provider):} This category includes repositories that aim to create and distribute a generation tool for the downstream community. For example, \kn{mobile} provides a package \kn{gobind} to generate the necessary bindings to call Go functions from programming languages frequently used on mobile platforms, like Java and Objective-C. }

%\del{Among these open-source projects, the project with the most frequent use of CGO function calls is \href{https://github.com/go-gl/gl}{\kn{gl}}, which is a Go binding for OpenGL. It contains 19 C-to-Go calls and 39755 Go-to-C calls, for a total of 39774. The reason why this project has so many CGO references is that it implements a binding for different versions of OpenGL, and each version contains thousands of call sites. Through in-depth manual analysis, we find that most of the use of CGO in project \kn{gl} is to wrap the library functions in OpenGL into Go functions through a wrapper function. As the example shown in \autoref{code:rq1:example}, writing this wrapper function is tedious, mechanical, and repetitive. Therefore, it is worth noting that the binding code in \kn{gl} is generated by a semi-automatic generator named \href{https://github.com/go-gl/glow}{\kn{glow}}. \kn{glow} is a dedicated Go binding generator for OpenGL, which parses the OpenGL XML API registry and the EGL XML API registry to produce a machine-generated CGO bridge between Go functions and native OpenGL functions. This highlights the feasibility of generating bridging code between Go and C via automated tools. }

\finding{1 to RQ1}{About 11.3\% of Go projects utilize CGO, highlighting its considerable presence within the developer community. CGO usage tends to be \cjbr{concentrated}{clustered}, often within a single package or a handful of files in a project. \cjbr{It is feasible to automatically generate CGO bridging code according to a certain configuration description.}{To avoid repetitive and tedious work, approximately 9.6\% of these CGO projects employ automated tools. This automation ranges from generating comprehensive API bindings to handling specific tasks, highlighting a pragmatic strategy to enhance development productivity and reliability.}}

\begin{table}[htb]
\centering
\caption{
    \cjb{Automated CGO bridging code generation tools used in CGO projects. 
    The \textbf{Type} column indicates the automation scope: \textit{Full} generates the complete binding interface; \textit{Partial} automates specific helper tasks; \textit{Provider} indicates the project itself is a generator tool for the downstream community. 
    The \textbf{Tool} column lists the used tool, where \textit{Custom} denotes a project-specific one.} 
}
\label{tab:cgo-automation-tools}
\scalebox{0.8}{
\begin{tabular}{llll}
\toprule

\cjb{\textbf{Repository}} & \cjb{\textbf{Tool}} & \cjb{\textbf{Type}} & \cjb{\textbf{Notes}} \\ 
\midrule

\cjb{\kn{gl}, \kn{glfw}} & \cjb{\href{https://github.com/go-gl/glow}{\kn{glow}}} & \cjb{Full} & \cjb{Generate bindings from configuration.} \\
\cjb{\kn{nuklear}} & \cjb{\href{https://github.com/xlab/c-for-go}{\kn{c-for-go}}} & \cjb{Full} & \cjb{Generate bindings from configuration.} \\
\cjb{\kn{qt}, \kn{pan-light}} & \cjb{Custom} & \cjb{Full} & \cjb{Template-based binding generation for QT.} \\
\cjb{\kn{macdriver}} & \cjb{Custom} & \cjb{Full} & \cjb{Generate bindings for Apple frameworks.} \\
\cjb{\kn{beats}} & \cjb{Custom} & \cjb{Partial} & \cjb{Map C error codes to Go error types.} \\
\cjb{\kn{confluent-kafka-go}} & \cjb{Custom} & \cjb{Partial} & \cjb{Map C error codes to Go error types.} \\
\cjb{\kn{cockroach}} & \cjb{Custom} & \cjb{Partial} & \cjb{Generate CGO build directives (preambles).} \\
\cjb{\kn{mobile}} & \cjb{\href{https://pkg.go.dev/golang.org/x/mobile/cmd/gobind}{\kn{gobind}}} & \cjb{Provider} & \cjb{Generate bindings for Java/Objective-C} \\

\bottomrule
\end{tabular}
}
\end{table}

\mysect{Purposes of Using CGO Beyond Binding.} CGO is commonly used as FFI to create Go bindings, but our analysis of the top 20 most-starred open-source CGO projects (\autoref{tab:top20repo}) uncovers a broader range of purposes. These projects span diverse domains such as containers, web servers, blockchain, and databases. %They are \href{https://github.com/moby/moby}{moby},\href{https://github.com/flannel-io/flannel}{flannel},\href{https://github.com/google/cadvisor}{cadvisor},\href{https://github.com/containers/podman}{podman}, \href{https://github.com/ehang-io/nps}{nps},\href{https://github.com/gravitational/teleport}{teleport},\href{https://github.com/getlantern/lantern}{lantern}, \href{https://github.com/gofiber/fiber}{fiber},\href{https://github.com/nats-io/nats-server}{nats-server}, \href{https://github.com/cockroachdb/cockroach}{cockroach},\href{https://github.com/dgraph-io/dgraph}{dgraph},\href{https://github.com/CodisLabs/codis}{codis}, {rclone},\href{https://github.com/cloudreve/Cloudreve}{Cloudreve}, \href{https://github.com/ethereum/go-ethereum}{go-ethereum}, \href{https://github.com/grafana/loki}{loki}, \href{https://github.com/fyne-io/fyne}{fyne}, \href{https://github.com/elastic/beats}{beats},\href{https://github.com/peterq/pan-light}{pan-light} and \href{https://github.com/prometheus/node_exporter}{node\_exporter}. 
Two authors independently annotated the modules in these repositories using the method described in \S \ref{sec:qualitative}. During this process, they recorded information like module paths, commit messages introducing CGO, CGO-related code comments, and the primary C libraries and header files used.

\input{top20repo}

% \del{We analyze these projects by module and record the following information in turn: }
% \begin{itemize}
%     \item %We record 
%     the path of the CGO-related module to be analyzed,
%     \item %We record 
%     the commit URL that introduces CGO,
%     \item %We record 
%     the C headers or libraries that are mainly used by the module,
%     \item some notes to help better understand.
% \end{itemize}
% \del{We then try to summarize the purposes of using CGO by analyzing CGO-related code comments, messages of commits that introduce CGO, and C headers or libraries used by the module.}

% The top 20 projects are \href{https://github.com/moby/moby}{moby},\href{https://github.com/flannel-io/flannel}{flannel},\href{https://github.com/google/cadvisor}{cadvisor},\href{https://github.com/containers/podman}{podman}, \href{https://github.com/ehang-io/nps}{nps},\href{https://github.com/gravitational/teleport}{teleport},\href{https://github.com/getlantern/lantern}{lantern}, \href{https://github.com/gofiber/fiber}{fiber},\href{https://github.com/nats-io/nats-server}{nats-server}, \href{https://github.com/cockroachdb/cockroach}{cockroach},\href{https://github.com/dgraph-io/dgraph}{dgraph},\href{https://github.com/CodisLabs/codis}{codis}, {rclone},\href{https://github.com/cloudreve/Cloudreve}{Cloudreve}, \href{https://github.com/ethereum/go-ethereum}{go-ethereum}, \href{https://github.com/grafana/loki}{loki}, \href{https://github.com/fyne-io/fyne}{fyne}, \href{https://github.com/elastic/beats}{beats},\href{https://github.com/peterq/pan-light}{pan-light} and \href{https://github.com/prometheus/node_exporter}{node\_exporter}.

% \begin{itemize}
%     \item First, we record the path of the module to be analyzed.
%     \item second
% \end{itemize}
% \input{sections/top20repo}

From these top 20 projects, we get 56 CGO-related modules and find that developers use CGO for the following 3 main purposes beyond binding:

    \kw{1) For getting system information and system control.}
    Our analysis indicates that 32 of the 56 modules are related to obtaining system information and control. These modules typically utilize headers with \codet{sys/} or \codet{linux/} on Linux, and \codet{mach/} on Mac, using APIs declared in these headers for tasks like getting CPU or memory status, system configuration and system control. For instance, in the \kn{quota} module of \href{https://github.com/moby/moby/tree/797b974cb90e32c7efe9e324d7459122c3f6b7b0}{\kn{moby}}, developers use \codet{linux/quota.h} and \codet{linux/dqblk\_xfs.h} for XFS quota control on Linux. In the \kn{internal/gopsutil/cpu} module of \href{https://github.com/gofiber/fiber/tree/5f1acd3c562f978318d07065153abbffdd6d4813}{\kn{fiber}} , they use header files like \codet{mach/processor\_info.h} to get CPU status on Mac. The prevalence of CGO for these tasks likely stems from the fact that Unix-like OS kernels are C-based and offer comprehensive C APIs for interaction.
    % \del{According to our analysis, 32 out of the 56 modules are about getting system information and system control. These modules usually use headers prefixed with \codet{sys/} or \codet{linux/} on Unix and \codet{mach/} on Mac. They use APIs declared in the headers to get system information such as CPU or memory status, and system configuration. They also use the APIs to control the system, such as making system calls. For example, in module \kn{quota} of project \href{https://github.com/moby/moby/tree/797b974cb90e32c7efe9e324d7459122c3f6b7b0}{\kn{moby}}, developers use \codet{linux/quota.h} and \codet{linux/dqblk\_xfs.h} for XFS quota control on Linux. In module \kn{internal/gopsutil/cpu} of \href{https://github.com/gofiber/fiber/tree/5f1acd3c562f978318d07065153abbffdd6d4813}{\kn{fiber}}, developers use headers such as \codet{mach/processor\_info.h} to obtain CPU status on Mac. A possible explanation for the high frequency of using CGO to get system information and control is that modern Unix-like operating system kernels are mainly written in C and provide a complete and efficient set of C APIs for interaction. }
    
    \kw{2) For reusing some functionalities.} 22 modules out of 56 leverage mature C libraries for specific functionalities. For example, in module \kn{pkg/ccl/gssapiccl} of \href{https://github.com/cockroachdb/cockroach/tree/893643b63ea0b1cfa4888c6b73b5c68a9c100c3a}{\kn{cockroach}}, developers use \kn{libkrb5} to add GSS authentication support on Linux. \kn{libkrb5} contains libraries for MIT Kerberos~\citep{MIT-Kerberos}, a system for authenticating users and services on a network. It is a trusted third party service with only official C implementation. In module \kn{perf} of \href{https://github.com/google/cadvisor/tree/6acde093f1acbf315025c4bf249d3c72de078ea2}{\kn{cadvisor}}, developers use \kn{libpfm}~\citep{libpfm} to measure perf events. \kn{libpfm} is a library to develop monitoring tools, and there is no similar Go library.
    %\item 
    
    \kw{3) For unlocking better performance.} Since C is a high-performance language, some Go developers use C to explore better performance. 4 out of the 56 modules mention in their commit messages, code comments, or README that CGO is introduced for performance optimization. For example, in module \kn{crypto/secp256k1} of \href{https://github.com/ethereum/go-ethereum/tree/d3e3a460ec947c9e1e963d1a35f887d95f23f99d}{\kn{go-ethereum}}, developers use \kn{libsecp256k1} for EC operations. \kn{libsecp256k1} is an optimized C library for EC operations on curve secp256k1 written by \kn{go-ethereum} developers. In module \kn{backend/udp} of  \href{https://github.com/flannel-io/flannel/tree/0083735ef9a28fbd5c8bc05f207bc17dfa176514}{\kn{flannel}}, developers use C to achieve faster proxy loop.

Apart from the 3 main purposes mentioned above, some other miscellaneous purposes do not account for much. For example, in module \kn{librclone} of \href{https://github.com/rclone/rclone/tree/c32d5dd1f38d3fe6a3727bd2cf683d57ffc008ad}{\kn{rclone}}, developers use CGO to export \kn{rclone} APIs for use in C.

\finding{2 to RQ1}{Developers primarily utilize CGO beyond binding for tasks such as obtaining system information, system control, implementing functionalities using other C libraries, and optimizing performance. This indicates C's potential advantages over Go in these areas. }

%% file: CGO-RQ1.tex
\begin{figure}[!htp]
    \centering
    \includegraphics[width=0.6\linewidth]{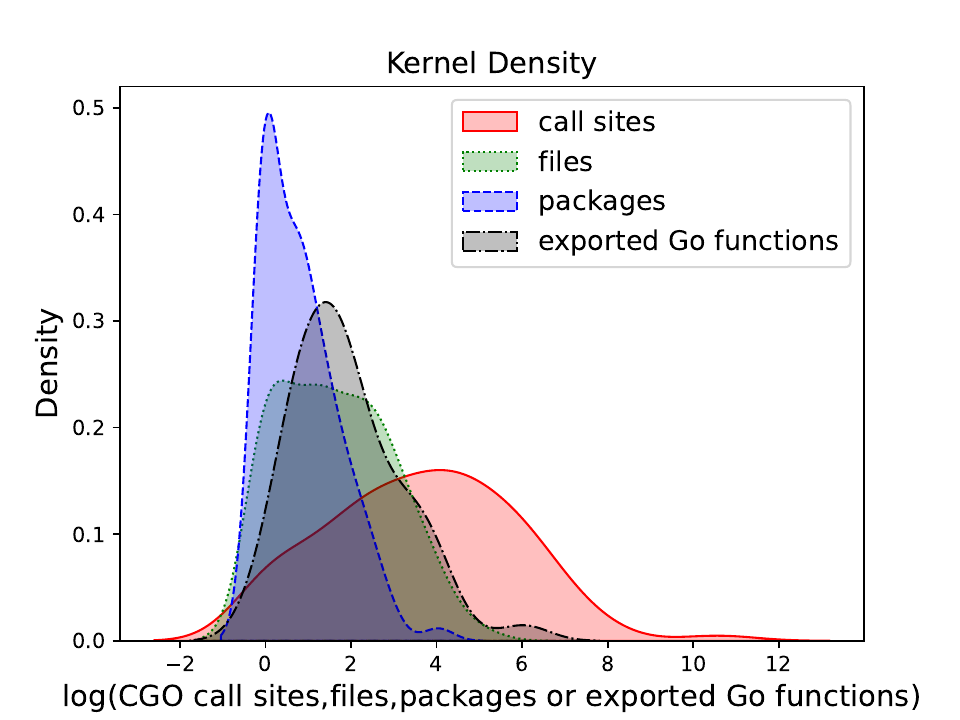}
    \caption{Kernel density of CGO usage at four levels}
    \label{rq1:kde}
\end{figure}

%% file: top20repo.tex
\begin{table}[ht]
\small
\centering
\caption{The top 20 most starred repositories using CGO}
\label{tab:top20repo}
\begin{tabular}{l|l|l}
\hline
\textbf{Domain} & \textbf{Repos} & \textbf{Avg Stars} \\ \hline
Container & \href{https://github.com/moby/moby}{moby},\href{https://github.com/flannel-io/flannel}{flannel},\href{https://github.com/google/cadvisor}{cadvisor},\href{https://github.com/containers/podman}{podman} & 22.8K \\ \hline
Proxy & \href{https://github.com/ehang-io/nps}{nps},\href{https://github.com/gravitational/teleport}{teleport},\href{https://github.com/getlantern/lantern}{lantern} & 12.7K \\ \hline
Web Server & \href{https://github.com/gofiber/fiber}{fiber},\href{https://github.com/nats-io/nats-server}{nats-server} & 12.8K \\ \hline
Database & \href{https://github.com/cockroachdb/cockroach}{cockroach},\href{https://github.com/dgraph-io/dgraph}{dgraph},\href{https://github.com/CodisLabs/codis}{codis} & 17.0K\\ \hline
Cloud Storage & \href{https://github.com/rclone/rclone}{rclone},\href{https://github.com/cloudreve/Cloudreve}{Cloudreve} & 19.8K\\ \hline
Blockchain & \href{https://github.com/ethereum/go-ethereum}{go-ethereum} & 32.6K \\ \hline
Log System & \href{https://github.com/grafana/loki}{loki}  & 13.9K \\ \hline
GUI & \href{https://github.com/fyne-io/fyne}{fyne} & 14.2K \\ \hline
Misc & \href{https://github.com/elastic/beats}{beats},\href{https://github.com/peterq/pan-light}{pan-light},\href{https://github.com/prometheus/node_exporter}{node\_exporter}  & 9.5K \\ \hline
\end{tabular}
\end{table}

%% file: rq2.tex
\subsection{RQ2: Most Commonly Used CGO Patterns}
\label{sec:empirical:rq2}
We manually inspect the top 20 most starred CGO repositories at the file level to understand commonly used CGO patterns and their applications. \cjb{Specifically, we analyze and label all 101 files containing CGO code within these 20 repositories to ensure a comprehensive view of the patterns in mature, well-regarded projects.}
\remv{Utilizing the method outlined in \mbox{\S \ref{sec:qualitative}}, we analyze CGO-related code in each repository and label CGO-related files.} Our analysis reveals 15 distinct patterns across 101 files, organized into 6 major categories. \cjb{This classification was derived using the systematic qualitative method described in \S \ref{sec:qualitative}. Specifically, two authors independently applied an open card-sorting approach, grouping similar CGO usages and assigning descriptive labels. The final categories were established through iterative discussion and reconciliation of their findings.} \remv{These classifications have been jointly discussed and agreed upon by two authors.}

In \autoref{tab:patterns}, we present 15 patterns alongside their corresponding categories, descriptions, and frequencies within the 101 files. For instance, considering the \kn{CType} pattern, it appears in 63 files, resulting in a frequency of 62.38\%. Notably, two patterns span multiple categories: \kn{Unsafe} and \kn{TypCast}. This occurs because developers may use the \codet{unsafe} package either for interfacing with C or for enhancing performance. Similarly, \kn{TypCast} may be combined with \kn{Unsafe} for better performance. Further elaboration on each category and pattern is provided below.

\input{pattern.tex}

\mysect{Communication without using \kn{Unsafe}.} Our findings emphasize the prevalence of communication patterns between Go and C in CGO usage. \autoref{code:pattern:com} lists six examples of typical \kn{Communication} patterns(\kn{CP1}$\sim$\kn{CP6}), corresponding to patterns \kn{CType}, \kn{CVar}, \kn{Export} and \kn{Malloc}. Direct communication requires declaring essential variables using C types (\kn{CP1}), creating Go types based on C types (\kn{CP2}), obtaining information via C macros (\kn{CP3}) or constants (\kn{CP4}), exporting Go functions for C usage (\kn{CP5}), and manually managing memory (\kn{CP6}) due to C's lack of automatic memory management.
%\del{For direct communication, developers need to declare some necessary variables using C types for later use in C~(\kn{CP1}), declare new Go types using C types for later use in Go~(\kn{CP2}), or get some necessary information via C macros~(\kn{CP3}) or constants~(\kn{CP4}). To interact with Go on the C side, developers need to export some Go functions for use in C. For example, using function \kn{secp256k1GoPani} declared in Go~(\kn{CP5}) can turn a C error into a recoverable Go panic. Besides, since C has no automatic memory management, developers may sometimes need to manually allocate and free memory to prepare the arguments for later CGO calls~(\kn{CP6}). }

\begin{center}
\begin{minipage}[c]{0.8\textwidth}

\begin{lstlisting}[
    style       =   Go
]
//CP1 from moby#797b97: declare variables using C types
var args C.struct_btrfs_ioctl_vol_args_v2 // CType
C.set_name_btrfs_ioctl_vol_args_v2(&args, cs)
//CP2 from fiber#5f1acd: declare Go types using C types
type Timespec C.struct_timespec // CType
//CP3 from moby#797b97: get C library version via C macro
return C.BTRFS_LIB_VERSION // CVar
//CP4 from fiber#5f1acd: declare Go constants using C
const sizeofLongLong = C.sizeof_longlong // CVar
//CP5 from go-ethereum#d3e3a4: export Go function
//export secp256k1GoPanicError // Export
func secp256k1GoPanicError(msg *C.char, ...) {panic()}
// CP6 from moby#797b97: allocate buffer for use in C
buffer := C.malloc(128) // Malloc
defer C.free(buffer) 
C.dm_task_get_driver_version(..., (*C.char)(buffer))
\end{lstlisting}
\captionof{figure}{\myinsert{ \kn{Communication} pattern examples without \kn{Unsafe}}}
\label{code:pattern:com}
\end{minipage}
\end{center}

\mysect{Unsafe usages for communication or performance.} \kn{Unsafe} is a CGO pattern using the \codet{unsafe} package, affecting both \kn{Communication} and \kn{Performance} categories. It bypasses Go's type safety, potentially compromising portability and compatibility with the ``Go 1''~\citep{unsafepackage, go1compat}. This pattern is often paired with patterns like \kn{TypCast} and \kn{StdCast}, which involve type casting between Go and C. \kn{StdCast} specifically utilizes built-in functions like \codet{C.CString}. Examples of \kn{Unsafe} combined with \kn{TypCast} or \kn{StdCast} are shown in \autoref{code:pattern:unsafe} (\kn{UP1}$\sim$\kn{UP4}).

In \kn{Unsafe} usage under \kn{Communication}, developers utilize \codet{unsafe.Pointer} to prepare arguments for C code. Certain C functions feature parameters of type \codet{void *}, which isn't directly available in Go, necessitating the conversion of Go pointers using \codet{unsafe.Pointer}. For example, in \kn{UP1}, to invoke the C function \codet{free} with a \codet{void *} parameter, developers convert \codet{cs} (of type \codet{char *}) to \codet{unsafe.Pointer}.

\begin{center}
\begin{minipage}[b]{0.8\textwidth}
\begin{lstlisting}[
    style       =   Go,
]
//UP1 from moby#797b97:
//use built-in functions to perform type casting, 
//then cast a Go pointer to unsafe.Pointer to call C.free
var cs = C.CString(q.backingFsBlockDev)
defer C.free(unsafe.Pointer(cs))
//UP2 from go-ethereum#d3e3a4:convert Go slice to C array
sigdata := (*C.uchar)(unsafe.Pointer(&signature[0]))
//UP3 from cockroach#893643: convert C array to Go slice
func cToUnsafeGoBytes(data *C.char, ...) []byte {
	return (*[Len]byte)(unsafe.Pointer(data))[:x:x] }
//UP4 from moby#797b97:cast between same layout types
Cinfo := C.struct_backport_dm_info{}
tmp := (*C.struct_dm_info)(unsafe.Pointer(&Cinfo))
\end{lstlisting}
\captionof{figure}{Examples combining \kn{Unsafe} with \kn{StdCast} or \kn{TypCast}}
\label{code:pattern:unsafe}
\end{minipage}
\end{center}

Go offers built-in functions for casting slices and arrays between Go and C, referred to as pattern \kn{StdCast}. These functions prioritize safety and compatibility but entail additional memory copy overhead. For instance, \codet{C.CString}, illustrated in \kn{UP1}, initially allocates memory in C and then copies the Go string into it. Besides the overhead of memory copying, developers must manually free the allocated C memory. 
 
\kn{Unsafe} also aids performance by bypassing memory copy overhead inherent in built-in functions for type casting between Go and C. This application falls under the \kn{Performance} category, the second most frequent category, encompassing \kn{Unsafe} and \kn{TypCast}. For instance, in \kn{UP2}, a Go byte slice can be efficiently cast to a C \codet{unsigned char} array by obtaining the address of the slice's first element and interpreting it as a \codet{C.uchar} pointer using \codet{unsafe.Pointer}. 
Similarly, a C pointer can be interpreted as a Go array pointer and then sliced (\kn{UP3}). Additionally, \codet{unsafe} allows bypassing type checking for efficient casting between types with matching memory layouts. In \kn{UP4}, \codet{struct\_dm\_info} and \codet{struct\_backport\_dm\_info} share identical memory layouts, enabling direct casting using \codet{unsafe.Pointer}.

\begin{center}
\begin{minipage}[c]{0.7\textwidth}
\begin{lstlisting}[
    style       =   Go,
]
// #cgo CFLAGS: -Wall -Werror -DDISABLE_JOIN_SHORTCUT
// #cgo LDFLAGS: -L/usr/lib -lperfstat
\end{lstlisting}
\captionof{figure}{\myinsert{\kn{Build} pattern examples}}
\label{code:pattern:build}
\end{minipage}
\end{center}

\mysect{CGO Build Process and Compilation.} The Go toolchain invokes the C compiler to process CGO-related C code. Developers utilize the \codet{\#cgo} directive in comments to specify compilation and linking options, such as include paths, link paths, library names, and macros, as depicted in \autoref{code:pattern:build}. These patterns are categorized under \kn{Build}. Notably, only \kn{cockroach} utilizes \codet{CXXFLAGS} and \codet{CPPFLAGS} because it relies on the C++ library \href{https://github.com/cockroachdb/geos}{\kn{GEOS}}. It's essential to note that CGO does not support C++, so developers must declare C++ functions in C format by adding \codet{extern "C"} to the C++ headers in order to use them.

\begin{center}
\begin{minipage}[c]{0.8\textwidth}
\begin{lstlisting}[
    style       =   Go,
]
// PP1 from moby#797b97: wrap a C function in Go
func dmTaskGetDriverVersionFct(task *cdmTask) string {
  buffer := C.malloc(128)
  defer C.free(buffer)
  res := C.dm_task_get_driver_version(task, ...)
  return C.GoString((*C.char)(buffer))
}
// PP2 from moby#797b97: embedded C code
/* static int get_priority(sd_journal *j, ...){
  rc = sd_journal_get_data(j,"PRIORITY",&data,&length);
  ... }*/
\end{lstlisting}
\captionof{figure}{\kn{Productivity} pattern examples}
\label{code:pattern:pro}
\end{minipage}
\end{center}

\mysect{Productivity-enhancing wrappers or embedded C code.} Category \kn{Productivity}, aimed at enhancing software development efficiency, also demonstrates high frequency. In \autoref{code:pattern:pro}, two typical \kn{Productivity} cases (\kn{PP1} and \kn{PP2}) are listed. Developers often encapsulate frequently used C functions in Go wrappers to streamline development. These wrappers not only invoke C functions but also manage the input and output contexts. In \kn{PP1}, C function \codet{dm\_task\_get\_driver\_version} is wrapped in a Go function, simplifying usage by managing type conversions and memory allocation. Moreover, developers may embed short but useful C functions in Go comments, 
\eg C function \codet{get\_priority} in \kn{PP2},
offering functionality not directly offered by the C library. %For instance, in \kn{PP2}, a concise C function to retrieve priority is embedded in a Go comment.

%\del{To make it easier to use in Go, the wrapper function also contains the context that converts \codet{task} to a proper C type, allocates memory for the C function, and finally converts the output to a proper Go string. Also, by embedding C code in Go comments, developers may implement some short but useful functions that are not directly provided by the used C library. For example, in \kn{PP2}, a short C function to get priority is embedded in the Go comment.}

\finding{3 to RQ2}{The predominant CGO pattern involves enabling direct communication between Go and C. Developers often utilize the \codet{unsafe} package to optimize CGO type conversions. To enhance productivity, they commonly encapsulate frequently used C functions and embed concise C functions within Go code as comments. Developers can learn mature CGO development practices from these CGO usage patterns identified in well-recognized projects.}

%% file: pattern.tex
\begin{table}[htb]
\centering\small
\caption{CGO usage patterns manually identified in 101 files}
\label{tab:patterns}
\scalebox{1}{
\begin{tabular}{l|l|l|l}
\hline
\bf Description & \bf Category & \bf Usage Pattern & \bf Frequency \\ \hline
\multirow{5}{*}{\begin{tabular}[c]{@{}l@{}}Communicate\\ between \\ Go and C\end{tabular}} & \multirow{5}{*}{Communication} & \texttt{CType} & 62.38\% \\ \cline{3-4} 
 &  & \texttt{CVar} & 47.52\% \\ \cline{3-4} 
 &  & \texttt{Export} & 7.92\% \\ \cline{3-4} 
 &  & \texttt{Malloc} & 5.94\% \\ \cline{3-4} 
 &  & \multirow{2}{*}{\texttt{Unsafe}} & \multirow{2}{*}{50.50\%} \\ \cline{1-2}
\multirow{2}{*}{\begin{tabular}[c]{@{}l@{}}For better \\ performance\end{tabular}} & \multirow{2}{*}{Performance} &  &  \\ \cline{3-4} 
 &  & \multirow{2}{*}{\texttt{TypCast}} & \multirow{2}{*}{48.51\%} \\ \cline{1-2}
\multirow{2}{*}{\begin{tabular}[c]{@{}l@{}}Perform \\ type casting\end{tabular}} & \multirow{2}{*}{Type Casting} &  &  \\ \cline{3-4} 
 &  & \texttt{StdCast} & 44.55\% \\ \hline
\multirow{5}{*}{\begin{tabular}[c]{@{}l@{}}To correctly \\ build from \\ source\end{tabular}} & \multirow{5}{*}{Build} & \texttt{LDFLAG} & 31.68\% \\ \cline{3-4} 
 &  & \texttt{CFLAG} & 12.87\% \\ \cline{3-4} 
 &  & \texttt{CPPFLAG} & 3.96\% \\ \cline{3-4} 
 &  & \texttt{PkgConfig} & 3.96\% \\ \cline{3-4} 
 &  & \texttt{CXXFLAG} & 2.97\% \\ \hline
\multirow{2}{*}{\begin{tabular}[c]{@{}l@{}}To increase\\ productivity\end{tabular}} & \multirow{2}{*}{Productivity} & \texttt{Wrapper} & 49.50\% \\ \cline{3-4} 
 &  & \texttt{Embedded} & 22.77\% \\ \hline
Other patterns & Other & \texttt{None} & 4.95\% \\ \hline
\end{tabular}
}
\end{table}

% 整体为通信
% wrapper, embed 提升生产力

% 通信：CType、CVar、Unsafe、Malloc

% 效率：unsafe、Malloc

% 构建选项：FLAG，PKGCONFIG

%% file: rq3.tex
\subsection{RQ3: Common Issues about CGO}
\label{sec:bug}

\begin{figure}[t]
    \centering
    \subfloat[%Issues - Categories - Labels of 
    \kn{1L1C}-class (issues with one label and one category)
    \label{fig:sankey:one2one}]{
        \includegraphics[width=0.30\linewidth]{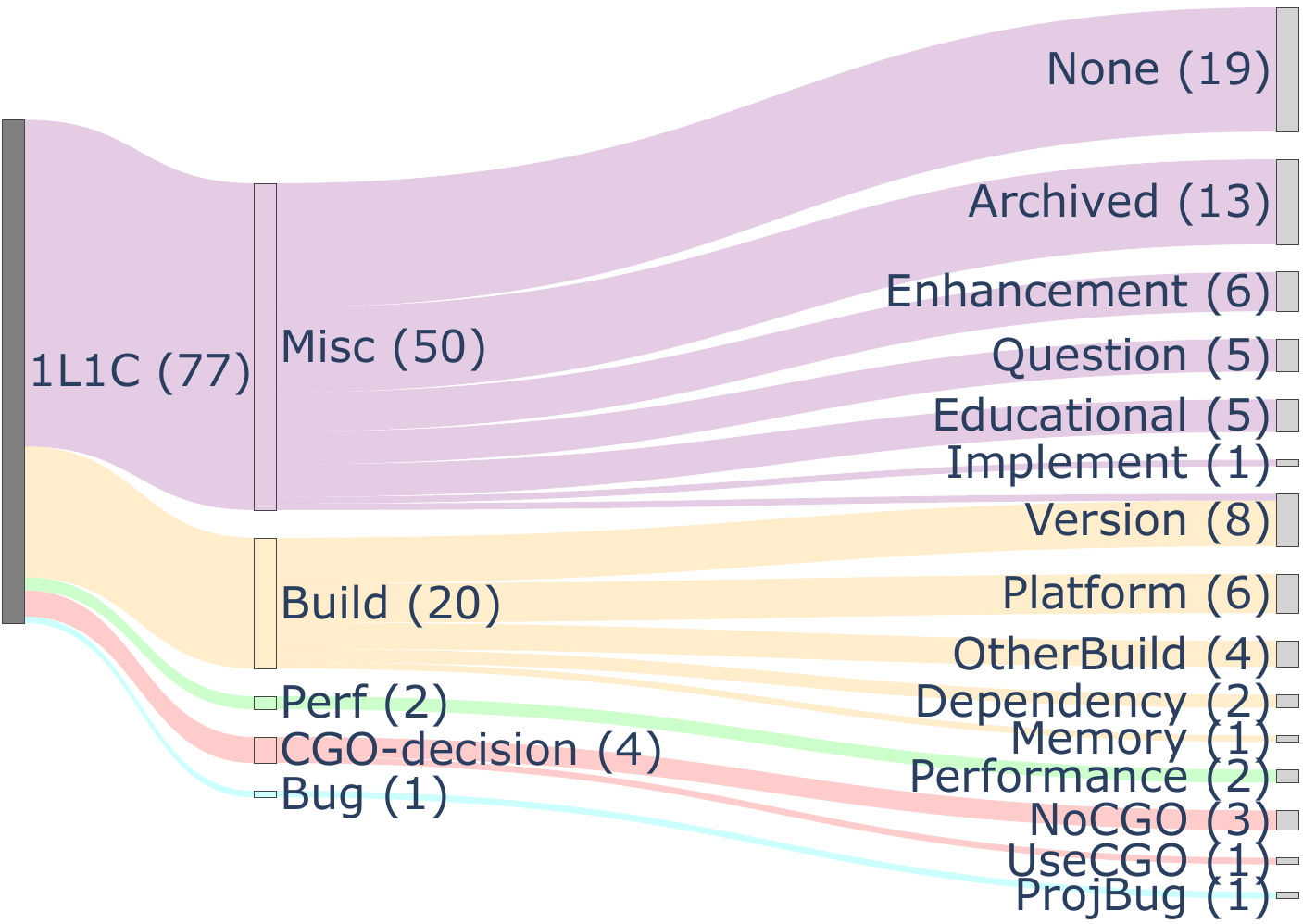}
    }
    \hspace{0.01\linewidth}
    \subfloat[%Issues - Categories - Labels of 
    \kn{NL1C}-class (issues with $N$ labels and one category)
    \label{fig:sankey:many2one}]{
        \includegraphics[width=0.30\linewidth]{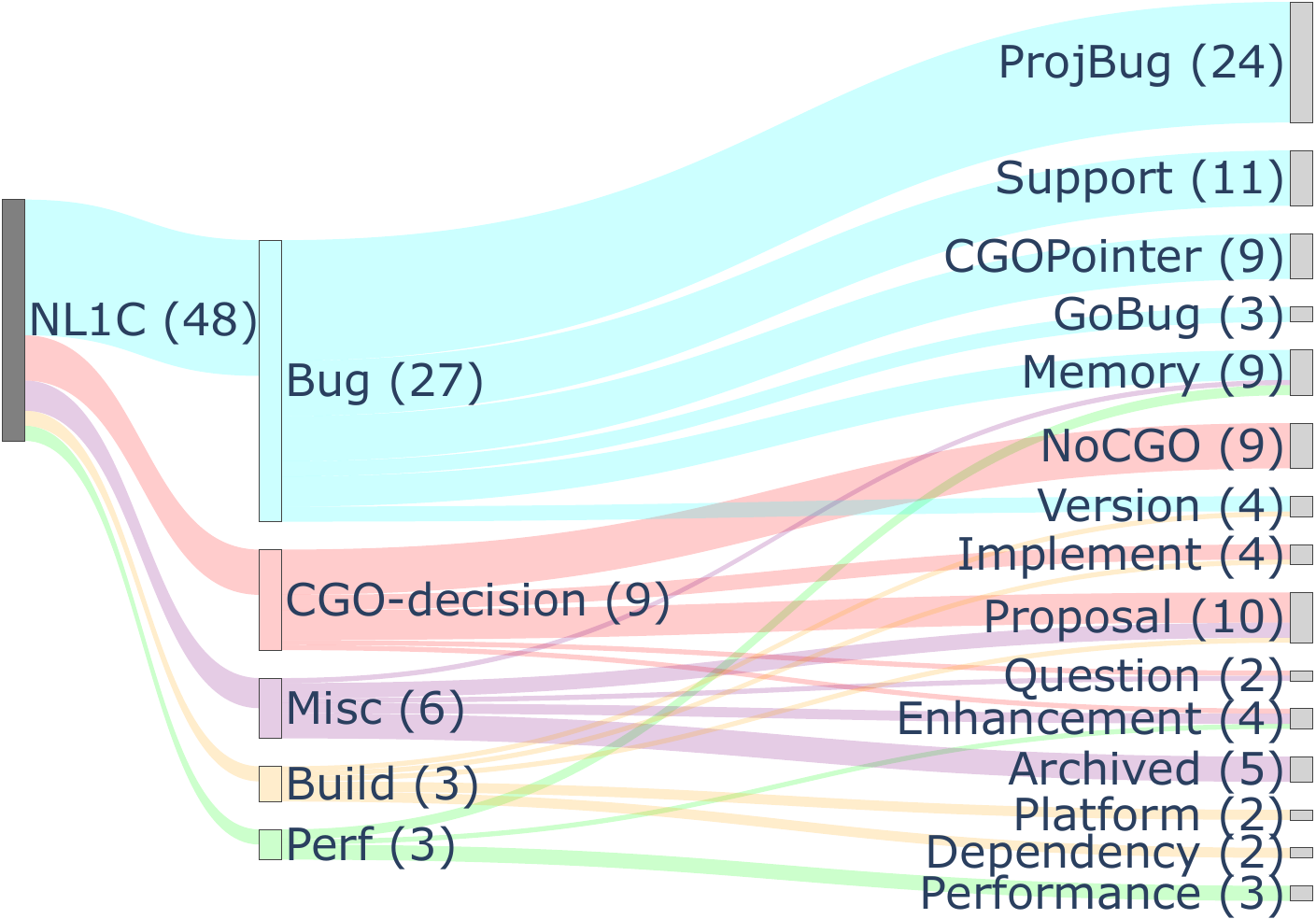}
    }
    \hspace{0.01\linewidth}
    \subfloat[%Issues - Categories - Labels of 
    \kn{Other}-class \label{fig:sankey:many2many}]{
        \includegraphics[width=0.30\linewidth]{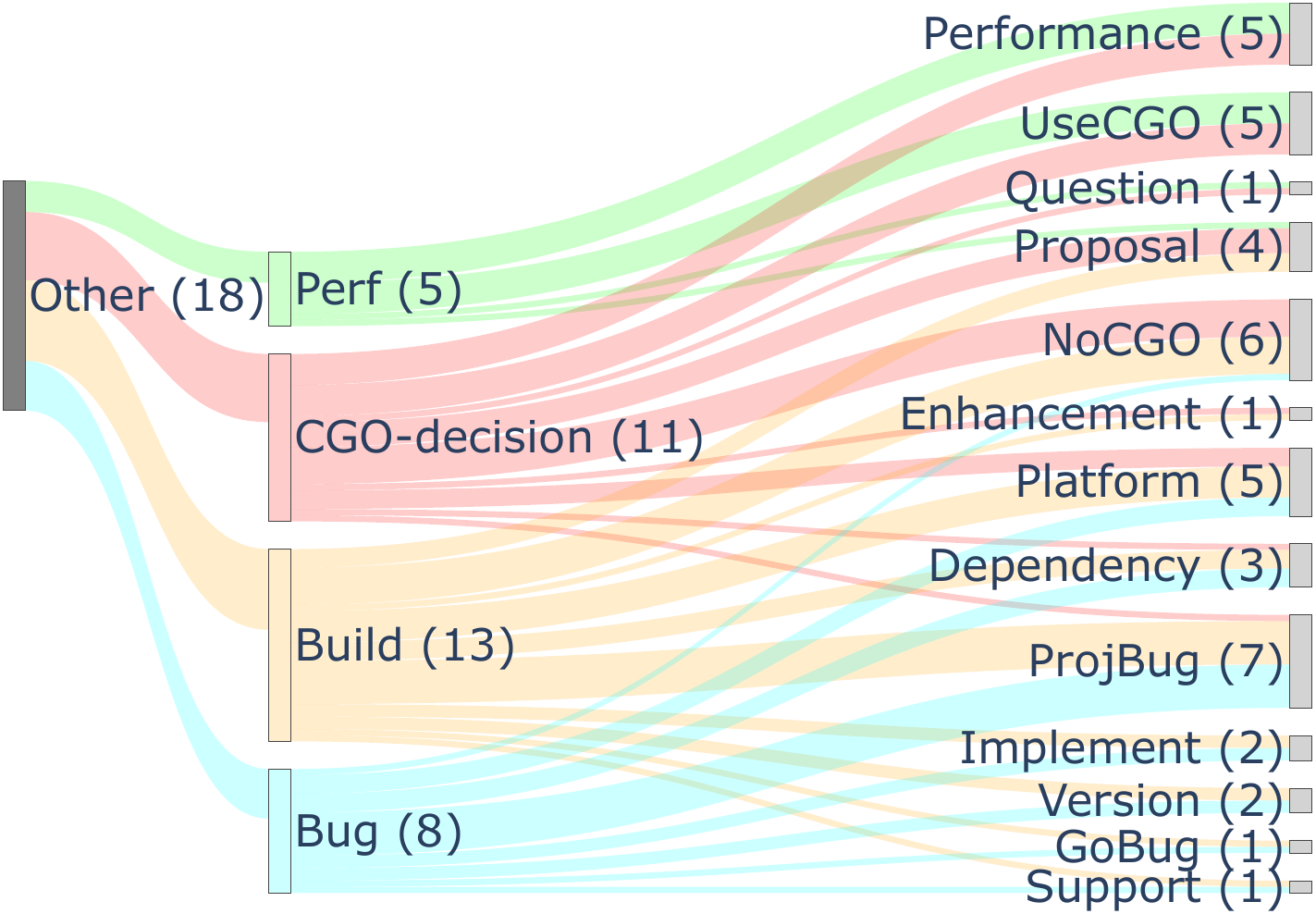}
    }
    \caption{The Sankey charts for three classes of Issues, Categories, and Labels. 
    Category \kn{Build} covers issues related to compiling and linking binaries. Category \kn{Bug} includes \kn{ProjBug} and \kn{GoBug} issues. Category \kn{CGO-decision} covers issues on whether to use CGO (labeled as \kn{UseCGO} and \kn{NoCGO}). Category \kn{Perf} includes performance-related issues with label \kn{Performance}. (Note: Since each issue may have multiple labels, the total count of issues for all labels within a category may exceed the number of issues in that category)}
    \label{fig:num_issue_label}
\end{figure}

We manually analyze CGO-related issues obtained in \S \ref{sec:mysys:issues} to identify common issues and highlight the critical ones. Each issue may be assigned one or multiple labels. Additionally, we categorize the labels into several categories. Finally, we obtained 19 labels, which are classified into five categories: \kn{Build}, \kn{Bug}, \kn{CGO-decision}, \kn{Perf}, and \kn{Misc}. Additionally, to better clarify the relationship between labels and categories, we further classify each issue into three mutually exclusive classes based on the number of labels and categories assigned to it: \kn{1L1C} (one label to one category), \kn{NL1C} (multiple ($N$) labels to one category), and \kn{Other} (multiple labels to multiple categories). The corresponding Sankey chart in \autoref{fig:num_issue_label} reveals the relations among issues, classes, categories and labels.  

For the 3 classes, 77 issues (53.8\%) fall into \kn{1L1C}, with the majority categorized under \kn{Misc} and \kn{Build}. The \kn{NL1C} class contains 48 issues (33.6\%), with the majority related to the \kn{Bug} category. The remaining 18 issues (12.6\%) fall into the \kn{Other} class, indicating that they have multiple labels and span multiple categories. For the 5 categories, the \kn{Build} category includes issues related to compilation and building and is mainly associated with labels \kn{Version}, \kn{Platform}, and \kn{Dependency}. The \kn{Bug} category focuses on issues related to CGO bugs, which may involve bugs within the project itself, the Go compilation toolchain, and so on. Key labels associated with this category include \kn{ProjBug}, \kn{GoBug}, \kn{Memory}, among others. The \kn{CGO-decision} category centers around discussions on whether to use CGO, with the main labels being \kn{NoCGO} and \kn{UseCGO}. The \kn{Perf} category addresses issues concerning the time and space performance of CGO, with \kn{Performance} being the primary label associated with these discussions. Finally, the \kn{Misc} category includes issues that are less relevant to this study and is linked to labels like \kn{Educational} and \kn{Question}. Below we will give a detailed analysis and explanation of these five categories of issues, describing their meanings, characteristics, and the insights they provide.

\mysect{Issues categorized as \kn{Build} account for the most}. The \kn{Build} category covers issues with building binaries from source, including dependencies, compilation, and linking. Out of 143 issues, 36 are related to \kn{Build}. Among these, 14 issues (labeled as \kn{Version}) stem from version discrepancies with the Go compiler or project dependencies. For instance, changes since Go 1.12 disallow mangled C names in packages using CGO, necessitating the use of documented CGO names like \codet{C.char} rather than generated mangled names like \codet{\_Ctype\_char}. This update leads to build failures due to mangled CGO names~\citep{go-flutter-issue261,beats-issue11054}. 
13 issues (labeled as \kn{Platform}) arise from target platforms or cross-compilation, where CGO compilation relies on specific C compilers and linked libraries for the target machine/OS, leading to non-reproducible builds. 7 issues  (labeled as \kn{Dependency}) involve difficulties in building dependent libraries with CGO, notably \kn{sqlite}~\citep{go-flutter-issue499,ko-issue52,cfssl-issue650} and \kn{glibc}~\citep{rclone-issue5090,cilium-issue5055}. Other concerns include compilation options and discussions about dropping CGO due to recurring build issues.

\autoref{fig:num_issue_label} further reveals that among the 36 issues in the \kn{Build} category, the majority(20 issues) are classified as \kn{1L1C}, indicating that these problems are predominantly driven by a single factor, such as version discrepancies, platform-specific challenges, or dependency issues. In contrast, 13 issues fall under the \kn{Other} class, reflecting more complex cases where multiple factors overlap. This distinction suggests that while many build issues can be addressed with targeted, straightforward solutions, a significant portion are multifaceted, requiring more comprehensive interventions. 

These issues about \kn{Build} highlight the importance of maintaining compatibility with Go compiler updates and adhering to CGO naming conventions to avoid build issues. Cross-compilation challenges stress the need for properly configured environments with suitable C compilers and libraries for target platforms. Effective dependency management, especially for libraries like \kn{sqlite} and \kn{glibc}, is crucial to mitigate build problems.
 
\mysect{Various bugs about CGO}. 
Bug-related issues fall under the \kn{Bug} category, totaling 36, which is the highest number, equal to that of the \kn{Build} category. Among them, 32 are labeled as \kn{ProjBug} and 4 as \kn{GoBug}.
\kn{ProjBug} indicates CGO-related bugs caused by either the issue author or the project where the issue was reported, while \kn{GoBug} signifies CGO-related bugs from defects in the Go compilation toolchain, independent of the analyzed project.

Of the issues labeled \kn{ProjBug}, 12 arise in repositories offering third-party tools for Go, such as checkers and tiny compilers. These issues aren't directly related to CGO usage but stem from inadequate support or false positives regarding CGO features in these tools. For example, in \href{https://github.com/dominikh/go-tools}{\kn{go-tools}}, issue\#133~\citep{go-tools-issue133} and \#673~\citep{go-tools-issue673} report false positives for CGO code. Similarly, in \href{https://github.com/tinygo-org/tinygo}{\kn{tinygo}}, issue\#1620~\citep{tinygo-issue1620} and \#1624~\citep{tinygo-issue1624} highlight insufficient CGO support. These instances highlight the complexity of CGO and the oversight of its comprehensive consideration by developers.

Notably, 9 issues labeled \kn{CGOPointer} among those categorized as \kn{Bug} share significant similarities. This label refers to issues raised by Go's CGO pointer check. These issues %~\cite{issue14210, issue18184, issue23391}
highlight unexpected runtime panics caused by \codet{\_cgoCheckPointer} inserted by the Go compilation toolchain. Some occurrences result from developers neglecting restrictions on passing pointers between Go and C, as seen in issue\#200~\citep{issue200} of repository \href{https://github.com/go-gl/glfw}{\kn{glfw}}. Others stem from design flaws in Go compilation toolchain (labeled as \kn{GoBug}), such as issue\#664~\citep{gopacket-issue664}, where type conversions involving struct fields trigger false positive runtime panics, confirmed in Go issue\#32970~\citep{issue32970}. Additionally, Go issue\#14210~\citep{issue14210} highlights CGO false positive panics on the address of slices from function calls. To confirm that these issues caused by the Go compilation toolchain are not isolated cases, and because issues related to Go compilation toolchain may not be raised in these projects but rather in the Go compilation toolchain repository itself, we search for related issues in the \kn{golang/go} repository, which contains the source code for the toolchain. The results reveal 67 issues potentially related to this problem, suggesting that this issue is not an anomaly but a widespread concern. Moreover, these issues span from 2016 to 2024, indicating that the problem may be a long-standing one, as evidenced by issues such as %issue\#14210, \#18184, \#23391, \#59954 and \#67333
\citep{issue18184, issue23391, issue59954, issue67333}.

Because these widespread issues labeled \kn{CGOPointer} constitute a significant proportion of CGO-related bugs and undermine the stability and usability of CGO, we identify them as \textbf{Critical}. These issues cause runtime panics that are difficult to diagnose and resolve, posing substantial challenges for developers and threatening the reliability of projects integrating CGO. They expose both developer misunderstanding and intrinsic limitations in CGO’s pointer handling mechanism, which is crucial for safe cross-language interaction. Addressing these problems is essential to improving CGO's reliability and developer experience. We will provide detailed insights into these critical issues and propose our improvement methods in \S \ref{sec:ptrchk}.

In addition to crashes caused by CGO pointer checks, using CGO can introduce memory-related bugs, labeled with \kn{Memory}. For instance, \kn{dgraph}'s issue\#361 addresses a CGO memory leak~\citep{draph-issue361}, while \kn{alertmanager}'s issue\#304 notes a segmentation fault due to Go calling a glibc function via CGO~\citep{alertmanager-issue304}. Due to C's lack of memory safety, its integration with Go, a memory-safe language, can lead to such issues.

CGO-related bug issues are the most frequent in the \kn{Bug} category, on par with CGO-related build issues, highlighting the significant challenges developers face. These issues, including runtime panics and segmentation faults, can undermine project stability and are difficult to diagnose and resolve, adding substantial complexity to development.

\mysect{Getting rid of CGO may achieve better reliability}.
Issues discussing the decision to use CGO are categorized under \kn{CGO-decision}. Issues labeled with \kn{UseCGO} indicate a decision to use CGO, while those labeled with \kn{NoCGO} indicate a decision not to use CGO. This category includes a total of 24 issues, with 18 labeled \kn{NoCGO} and 6 labeled \kn{UseCGO}. The majority of the discussions focus on not using CGO.
%Among the issues, 18 out of 143 are related to removing CGO, labeled as \kn{NoCGO}. 
For instance, \kn{concourse}'s issue\#4342~\citep{concourse-issue4342} and \kn{cilium}'s issue\#5055\remv{(Thomas Graf, 2018)}\cjb{~\citep{cilium-issue5055}} %~\citep{issue5055}
highlight that CGO introduces a dependency on glibc, limiting applications to non-glibc based platforms. Similarly, \kn{cosmos-sdk}'s issue\#4280~\citep{issue4280} proposes dropping CGO for reproducible builds, as pure Go programs allow for reproducibility through static compilation and linking, which CGO compromises. The prevalence of issues recommending against CGO due to its build challenges supports our earlier finding: CGO-related build issues make up the highest proportion of CGO-related issues. 

Those suggesting the use of CGO (labeled as \kn{UseCGO}) are mostly due to the potential for better performance. For example, \kn{rclone}' issue\#4944~\citep{rcloneissue4944} recommends using CGO to list all files, which could result in several times the performance improvement. 

From the issues in this category, it is evident that while some developers use CGO for better performance, the majority believe removing CGO enhances reliability and simplicity. Avoiding CGO will eliminate platform dependencies, reduce build issues, and lower the risk of runtime errors, ultimately resulting in more robust and maintainable software.

\mysect{Space and time performance}. 
Performance-related issues are labeled as \kn{Performance} and categorized as \kn{Perf}. For example, cockroach's issue\#38116~\citep{cockroach-issue38116} compares MVCCScan and lower-level CGO iterator interface performance. In cockroach's issue\#10050~\citep{cockroach-issue10050}, developers discuss excessive CGO memory usage. Similarly, in flannel's issue\#368~\citep{flannel-issue368}, developers mention using CGO for performance gains.

These performance-related discussions highlight the trade-offs developers face with CGO. While some argue that CGO can improve performance in specific scenarios, others point out the additional overhead and memory usage it introduces. The analysis suggests that while CGO may be beneficial for certain performance-critical operations, its impact on memory management and system efficiency can be a significant downside. This emphasizes the need for \kw{carefully weighing performance gains against the potential costs}.

\mysect{Miscellaneous issues}.
Other labels such as \kn{Educational}, \kn{Question}, and \kn{None}, which fall under the \kn{Misc} category are either irrelevant to CGO or occupy a small proportion. Hence, we won't discuss them here.
\del{Other labels like \kn{Educational}, \kn{Question} and \kn{None} are either irrelevant to CGO \mydel{vulnerabilities} \myinsert{bugs} or account for only a small amount. So we don't discuss them here. }  

\finding{4 to RQ3}{Integrating CGO can increase project complexity and instability, leading to build issues and CGO-related bugs. While build issues often stem from dependency management and platform-specific configurations, and memory-related bugs arise from the unsafe nature of C's memory model, the most \textbf{Critical} issues lie in CGO pointer checks, which are related to defects in the Go compilation toolchain. These checks often result in false positives and unexpected runtime panics, as evidenced by a notable subset of issues labeled \kn{CGOPointer}. These issues are particularly concerning, as they undermine project stability and reliability, necessitating further investigation to address the root causes. For projects relying on non-essential C libraries, deprecating CGO can enhance reliability. }

%% file: ptrchk.tex
\section{Study on RQ4 - Critical CGO Pointer Check Issues} 
\label{sec:ptrchk}
%https://stackoverflow.com/questions/56027718/how-to-pass-a-go-pointer-to-cgo
%安全性，互相cast
%讲去掉一些没用的cgocheck，去掉后也是安全的，去掉后还减少误报的可能
%目前太保守，默认unsafe是不安全，但是cast到unsafe是安全的，因为我们确定cast到unsafe的指针的具体类型和分配位置
%一个目前可能误报的例子(slice+括号)
This section provides detailed insights into the critical \kn{CGOPointer} issues identified in RQ3 (\S \ref{sec:bug}). We first analyze the root causes of the issues in \S \ref{sec:ptrchk:issue}. Then in \S \ref{sec:ptrchk:allsol} and \S \ref{sec:ptreval}, we propose and evaluate our solutions, including a temporary but effective approach and a proposal for a permanent solution submitted to the Go team. Finally, in \S \ref{sec:ptrchk:discuss}, we discuss scenarios where such critical issues may arise.
%This section analyzes reasons \myR{of}{for} some issues mentioned in RQ3 and then presents our new strategy to insert function \codet{\_cgoCheckPointer}.  We answer RQ4 in this section. %\yunote{todo}

%\subsection{Potential Inadequacy of CGO's Pointer Check}
\subsection{The Root Cause}
\label{sec:ptrchk:issue}
To identify the root cause of the runtime crashes caused by the Go compilation toolchain in such critical issues, we analyze \kn{cmd/cgo}~\citep{gocmdcgo}, the component responsible for handling CGO within the toolchain. As mentioned in \S \ref{sec:bg:restriction}, Go imposes certain restrictions that developers must follow when passing pointers between Go and C using CGO. To ensure these restrictions are not violated and to maintain the correct execution of the Go runtime, \kn{cmd/cgo} inserts \codet{\_cgoCheckPointer} during CGO preprocessing for potentially suspicious pointers. At runtime, \codet{\_cgoCheckPointer} verifies compliance with these restrictions and triggers a panic if a violation is detected, terminating the program. However, there is a gap between current implementation of \kn{cmd/cgo} and the actual restrictions, causing \kn{cmd/cgo} to sometimes insert checks for valid pointers, which may lead to false positive panics. 
% 如我们在\S \ref{sec:bg:restriction}所述，Go imposes certain restrictions that developers must follow when passing pointers between Go and C using CGO. 为了确保上述约束不被违反，保证Go运行时正常运行，\kn{cmd/cgo}在CGO预处理时为可疑的指针插入\codet{\_cgoCheckPointer}。\codet{\_cgoCheckPointer}在运行时检查，如果发现违反限制，则触发panic，终止程序。然而，现有\kn{cmd/cgo}的实现和实际的约束之间存在gap，\kn{cmd/cgo}有时候会为合法的指针插入检查，并可能导致false positive panic. 具体来说

Specifically, current \kn{cmd/cgo} uses a fast yet conservative strategy to determine whether to insert \codet{\_cgoCheckPointer} for function formal parameters. It checks if the parameter's type may contain a pointer but does not track or infer the actual argument's type during CGO preprocessing.
%\del{Currently, the Go compiler adopts a relatively fast but conservative strategy of checking whether a formal parameter's type may contain a pointer to determine whether to insert a \codet{\_cgoCheckPointer} function call without tracking at compile time to infer the type of the actual argument.}
% This results in \codet{\_cgoCheckPointer} being inserted when the parameter type is an ambiguous pointer type.\yunote{DBY,CJB please check}
%. The Go compiler for now only checks the types of the formal parameters of a CGO call and does not track the actual types of the arguments to determine whether to insert a \codet{\_cgoCheckPointer}. The formal parameter type \codet{void *} in C corresponds to \codet{unsafe.Pointer} in Go. 
The \kn{cmd/cgo} translates \codet{void *} in C to \codet{unsafe.Pointer} in Go. When encountering an argument with a formal parameter type of \codet{unsafe.Pointer}, \kn{cmd/cgo} assumes conservatively that the memory pointed to by the argument may contain a pointer. Consequently, it inserts a \codet{\_cgoCheckPointer} function call. This behavior applies universally: if a C function called by Go has a \codet{void *} pointer in its formal parameter, a \codet{\_cgoCheckPointer} will always be inserted, regardless of the actual type pointed to by the argument.

% https://www.sobyte.net/post/2022-04/golang-memory-allocation/
% https://github.com/golang/go/blob/master/src/runtime/cgocall.go

\kn{mspan} is a fundamental memory management unit in Go responsible for managing memory pages. During runtime, function \codet{\_cgoCheckPointer} first examines the type of the incoming \codet{unsafe.Pointer} argument $p$. If $p$ points to an unknown type, the function then accesses the \kn{mspan} pointed to by $p$ and searches the entire \kn{mspan} to see if the Go memory contains a Go pointer. If such a pointer is found, a runtime panic is triggered.

\begin{center}
\begin{minipage}[htb]{0.85\textwidth}
\begin{lstlisting}[
    style       =   Go,
]
...
func main() {
	var a _Ctype_int = 1
	func() { _cgo0 := (unsafe.Pointer(&a)); _cgoCheckPointer(_cgo0, nil); _Cfunc_print_ptr(_cgo0); }() 
}
\end{lstlisting}
\captionof{figure}{\textbf{\autoref{code:cgo:go2c} after CGO preprocess}}
\label{code:cgo:unsafe-arg-pre}
\end{minipage}
\end{center}

\begin{center}
\begin{minipage}[c]{0.38\textwidth}
\begin{lstlisting}[style=Go]
var p *byte
p = &b[0]
C.testPtr(unsafe.Pointer(p))
\end{lstlisting}
\captionof{figure}{\textbf{Unrecognized address taken}}
\label{code:ptrchk:addrtak}
\end{minipage}
\hfill
\begin{minipage}[c]{0.55\textwidth}
\begin{lstlisting}[style=Go]
var p *byte
p = &b[0]
func() { _cgo0 := unsafe.Pointer(p); _cgoCheckPointer(_cgo0, nil); _Cfunc_testPtr(_cgo0); }()
\end{lstlisting}
\captionof{figure}{\textbf{\autoref{code:ptrchk:addrtak} after CGO preprocess}}
\label{code:ptrchk:addrtakpreprocess}
\end{minipage}
\end{center}

The above conservative strategy adopted by \kn{cmd/cgo} can lead to false positive runtime panics~\citep{issue18184, issue14210, issue32970, issue23391}. It may insert checks for valid pointers and trigger a runtime scan of the entire \kn{mspan}, even when the pointer itself does not violate any restrictions, such as when it points to a Go pointer. However, if the scanned \kn{mspan} contains any Go pointers, the runtime may mistakenly conclude that the checked pointer also points to a Go pointer, incorrectly identifying a violation and triggering a panic. Taking \autoref{code:cgo:go2c} as an example, a Go program passes an \codet{unsafe.Pointer} pointing to an \codet{int} variable \codet{a} to C. Though variable \codet{a} does not contain a pointer, \kn{cmd/cgo}'s conservative approach still inserts an call to \codet{\_cgoCheckPointer}  (as shown in \autoref{code:cgo:unsafe-arg-pre}). 
%\del{However, this conservative strategy of inserting \codet{\_cgoCheckPointer} at compile time may cause false positive runtime panics~\cite{issue18184, issue14210, issue32970, issue23391}. \autoref{code:cgo:go2c} shows a Go program that passes an \codet{unsafe.Pointer} which points to an \codet{int} variable \codet{a} to C. Variable \codet{a} obviously contains no pointer. Due to the conservative approach taken by \kn{cmd/cgo}, an unnecessary \codet{\_cgoCheckPointer} function  call is inserted during compilation (see \autoref{code:cgo:unsafe-arg-pre}). }
This increases the risk of a false positive runtime panic if the \kn{mspan} containing \codet{a} includes Go pointers. Furthermore, Go's issue\#18184~\citep{issue18184} highlights \kn{cmd/cgo}'s failure to recognize address-taking on a \codet{byte} slice element, as shown in \autoref{code:ptrchk:addrtak} and \autoref{code:ptrchk:addrtakpreprocess}. Current \kn{cmd/cgo} cannot determine if the \codet{unsafe.Pointer} points to a slice element and thus inserts a \codet{\_cgoCheckPointer} to check the entire \kn{mspan}, potentially leading to a false positive runtime panic.

All in all, the root cause behind these pointer-check-related issues is that the \kn{cmd/cgo} sometimes cannot figure out what \kw{type} \codet{unsafe.Pointer} actually points to. In such cases, \kn{cmd/cgo} inserts a pointer-check, and Go conservatively searches the whole \kn{mspan} during runtime in the check. To fundamentally address these issues, it is necessary to identify the type of object that \codet{unsafe.Pointer} points to, which requires incorporating type inference and pointer analysis passes in the compiler. However, integrating current \kn{cmd/cgo}, which handles CGO in the toolchain, with the type inference and pointer analysis passes of the Go compiler is a challenging task. It's because:
\begin{enumerate}
    \item Currently, the decision to insert \codet{\_cgoCheckPointer} occurs at the AST representation stage, which is unsuitable for pointer analysis.
    \item Currently, \kn{cmd/cgo} exists as a standalone tool within the Go compilation toolchain rather than being part of the Go compiler itself.
\end{enumerate}

Therefore, integrating \kn{cmd/cgo} into the compiler and enabling it to reuse the type inference and pointer analysis passes would require modifying the existing Go compilation process. This is a significant undertaking, and for broader applicability, it would require the involvement of the Go team. To address this, we propose a two-phase solution:
\begin{enumerate}
    \item Initially, implement a temporary but effective approach that can handle most scenarios.
    \item Submit a proposal to the Go team, hoping they will address the issue in the future.
\end{enumerate}

In \S \ref{sec:ptrchk:allsol}, we describe our temporary but effective approach as well as the proposal we submit. In \S \ref{sec:ptreval}, we evaluate the effectiveness of our approach.

%To figure out what type \codet{unsafe.Pointer} points to, we need a precise points-to analysis. 
% However, precise points-to analysis does not work with the current Go compiler. 
% \del{First, the time complexity of precise points-to analysis is $O(n^3)$, while Go prefers and is now taking a more lightweight algorithm~\cite{pearce2004pointer} with time complexity $O(n^2)$. }\myins{First, precise points-to analysis has a time complexity of $O(n^3)$~\cite{pearce2004pointer}, resulting in significant time overhead, which contradicts Go's ethos of rapid compilation.}
% \myins{Second, decisions regarding \codet{\_cgoCheckPointer} insertion are made earlier during AST (Abstract Syntax Tree) form operations, which are not suitable for points-to analysis at this stage.}
% If we use precise points-to analysis to determine whether to insert \codet{\_cgoCheckPointer}, we have to modify the current compilation process of Go. Due to the $O(n^3)$ time complexity of precise points-to-analysis and current compilation process, the compilation time of the Go compiler would \myins{substantially increase}\del{be much longer than before} and the work\myins{load} of modifying Go's current compilation process is expected to be \del{very heavy}\myins{extremely burdensome}.

\subsection{Solution}
\label{sec:ptrchk:allsol}

\subsubsection{A Temporary but Effective Solution}
\label{sec:ptrchk:sol}
% 重复了 Go is a garbage collected language, and the garbage collector needs to know the location of every pointer to Go memory. Because of this, there are such restrictions on passing pointers between Go and C \cite{cgopointerpass}. The Go compiler dynamically checks these restrictions during runtime through \codet{\_cgoCheckPointer}. The compiler inserts \codet{\_cgoCheckPointer} for any parameters that possibly have Go pointers during compile-time. 

%\autoref{code:cgo:not-unsafe-arg} shows a program that is very similar to \autoref{code:cgo:go2c}. The only difference is that the parameter type of \codet{print\_ptr} in \autoref{code:cgo:not-unsafe-arg} is \codet{int *}. \autoref{code:cgo:not-unsafe-arg-pre} is the form of \autoref{code:cgo:not-unsafe-arg} after CGO preprocess. Compared with \autoref{code:cgo:unsafe-arg-pre}, we can find that no \codet{\_cgoCheckPointer} functions are inserted. Using \codet{*C.int} escapes from the CGO pointer check, but Using \codet{unsafe.Pointer} doesn't, even though they contain the same pointer. 

\begin{center}
\begin{minipage}[c]{0.45\textwidth}
\begin{lstlisting}[
    style       =   Go,
    basicstyle  =   \footnotesize\ttfamily,
]
//void print_ptr(int *p);
func main() {
	var a C.int = 1
	C.print_ptr((&a))
}
\end{lstlisting}
\captionof{figure}{\textbf{Modified \autoref{code:cgo:go2c}}}
\label{code:cgo:not-unsafe-arg}
\end{minipage}
\hfill
\begin{minipage}[c]{0.45\textwidth}
\begin{lstlisting}[
    style       =   Go,
    basicstyle  =   \footnotesize\ttfamily,
]
...
func main() {
	var a _Ctype_int = 1
	( _Cfunc_print_ptr )((&a))
}
\end{lstlisting}
\captionof{figure}{\textbf{Preprocessed \autoref{code:cgo:not-unsafe-arg}}}
\label{code:cgo:not-unsafe-arg-pre}
\end{minipage}
\end{center}

We further analyze \autoref{code:cgo:go2c}. 
%\del{If we remove the \codet{unsafe.Pointer} wrapper of \codet{\&a} in line 10 and change the type of \codet{p} in line 3 as well as the C function parameter to \codet{int *}, we will get \autoref{code:cgo:not-unsafe-arg}.}
\autoref{code:cgo:not-unsafe-arg} is generated by removing the \codet{unsafe.Pointer} wrapper from \codet{\&a} in line 10 and changing the type of \codet{p} in line 3, along with the C function parameter, to \codet{int *}. \autoref{code:cgo:not-unsafe-arg-pre} shows what \autoref{code:cgo:not-unsafe-arg} looks like after CGO preprocess. Compared to \autoref{code:cgo:unsafe-arg-pre}, no \codet{\_cgoCheckPointer} is inserted. This shows that using \codet{*C.int} bypasses CGO pointer checks, unlike using \codet{unsafe.Pointer}, even when they refer to the same pointer. 
%\del{This means that using \codet{*C.int} will escape CGO pointer checks, but using \codet{unsafe.Pointer} will not, even if they act on the same pointer.}\del{ The reason is that Go is able to safely pass the exact pointer type to C. In fact, casting an exact pointer (like \codet{*C.int}) to \codet{unsafe.Pointer} is actually safe because Go can already pin the value. What is really dangerous is casting from an \codet{unsafe.Pointer}, because usually it is not clear what type is being pointed to.}
The distinction arises because Go safely passes exact pointer types to C. Casting an exact pointer (\codet{*C.int}) to \codet{unsafe.Pointer} is safe because Go can already pin the value accurately. Conversely, casting from an \codet{unsafe.Pointer} is risky because it's often unclear what type it points to. 
%\del{CGO assumes that every \codet{unsafe.Pointer} is potentially dangerous and conservatively inserts a \codet{\_cgoCheckPointer}. This partly explains the false positive panic. }
CGO treats every \codet{unsafe.Pointer} as potentially dangerous and thus conservatively inserts a pointer check, partly explaining the false positive panic.

\begin{figure}[htb]
    \centering  %图片全局居中
    \includegraphics[width=0.75\linewidth]{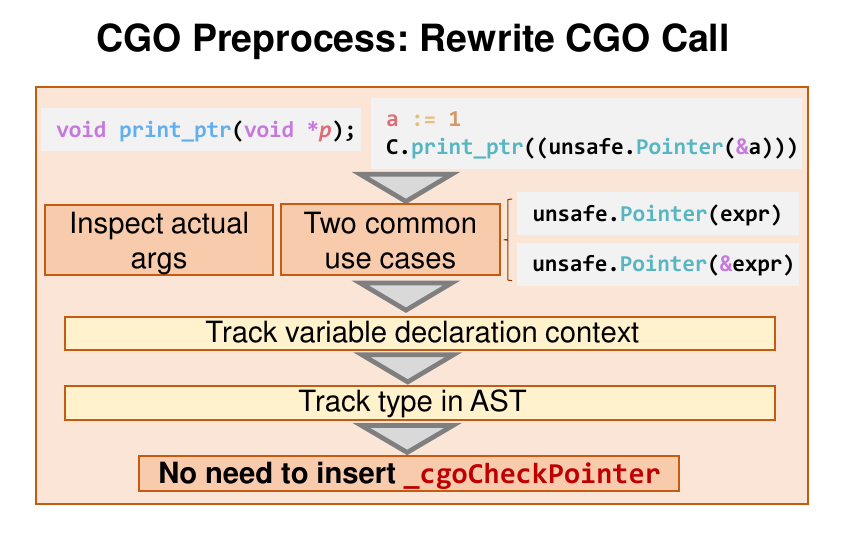}
    \caption{Architecture of our temporary method}
    \label{fig:ptrcheck}
\end{figure}

Based on the above observation, we propose a temporary but effective method: identify common and obvious safe cast to \codet{unsafe.Pointer} and avoid inserting \codet{\_cgoCheckPointer} on such pointers. We focus on two special but common cases shown in \autoref{fig:ptrcheck}: \codet{unsafe.Pointer(expr)} and \codet{unsafe.Pointer(\&expr)}. The two cases cast a Go pointer type directly to \codet{unsafe.Pointer} without any intermediate transformations before using this \codet{unsafe.Pointer}. In such cases, what \codet{unsafe.Pointer} really points to is \codet{expr}. We track the declaration of \codet{expr} to get its type. %If \codet{expr} is declared in a \codet{ValueSpec} node in Go's AST (like \codet{var expr type}) then its type is specified by \codet{type}\del{the variable declaration}. If \codet{expr} is declared in an \codet{AssignStmt} node\del{ in Go's AST, } (like \codet{expr := 1}), then its type is equal to the type of the corresponding rvalue. 
Specifically, We recursively inspect AST nodes related to \codet{expr} 
%including \kn{Ident, BasicLit, CompositeLit, CallExpr, SelectorExpr, UnaryExpr, IndexExpr},
to determine the type of \codet{expr}. If we cannot determine whether it is a safe cast, we default to Go's \myins{original} approach.

% \vspace{-0.1cm}

\subsubsection{Proposal for a permanent fix}
\label{sec:ptrchk:proposal}
% 我们的proposal介绍了该类issue的背景和原因，分析了现有CGO的不足并提出了解决方案。具体来说，我们的解决方案包括以下几步：
% 1.Integrating CGO as part of the Go compiler, rather than as a standalone tool.
% 2. Allowing CGO to reuse the compiler's analysis pass results, such as type inference and pointer analysis results. 
% By reusing these results, it can more accurately determine the actual type that unsafe.Pointer points to, thereby inserting _cgoCheckPointer more precisely and avoiding the aforementioned false positives. 
% 目前，我们的proposal被Go team group with 了一个老的，已经被接受，ongoing的proposal\cite{proposal16623}， 该proposal也是关于CGO的，是关于arrange to pass unmodified Go source files to compiler
% 我们的proposal和这个老issue被group with是因为我们计划的第一步：Integrating CGO as part of the Go compiler, rather than as a standalone tool. 与这个老issue有关。
% 目前，我们的proposal已经作为这个老issue的supplement，为其添加了额外consideration：
% 1. 要注意CGO接口中指针检查导致的false positve panic
% 2. 复用compiler中类型推断和指针分析的结果来从根本上解决这个问题
To further enhance CGO and permanent fix these issues, we submit a formal proposal~\citep{proposal70274} for the Go team. Our proposal introduces the background and causes of this class of issues, analyzes the limitations of the current CGO, and presents a solution. Specifically, our solution includes the following steps:
\begin{enumerate}
    \item Integrating CGO into the Go compiler rather than keeping it as a standalone tool.
    \item Allowing CGO to reuse the compiler's analysis pass results, such as type inference and pointer analysis results.
\end{enumerate}
By leveraging these results, CGO can more accurately determine the actual type that \codet{unsafe.Pointer} points to, enabling more precise insertion of \codet{\_cgoCheckPointer} and avoiding the aforementioned false positive crashes. 

Our proposal has been grouped by the Go team with a legacy, ongoing, but accepted proposal~\citep{proposal16623}, which focuses on arranging for unmodified Go source files to be passed to the compiler. The grouping stems from the connection between our first step, integrating CGO into the Go compiler, and the goals of the legacy proposal. %The grouping is due to the similarity between our first step, integrating CGO into the Go compiler and the objectives of the legacy proposal.
Currently, our proposal serves as a supplement to the legacy proposal, contributing additional considerations: pay attention to such pointer-check related crashes, and, reusing the compiler's type inference and pointer analysis results to fundamentally resolve the issue. This legacy accepted proposal has been around for a long time and requires the Go team to commit to refactoring parts of the Go compiler to enable CGO to reuse such passes. We hope that in the future, this accepted proposal will continue to make progress and ultimately address the current limitations of CGO, while also enhancing the overall robustness and security of the Go toolchain.

\subsection{Evaluation of our temporary approach}
\label{sec:ptreval}

\begin{center}
\begin{minipage}[c]{0.8\textwidth}
%\vspace{-0.1cm}
\begin{lstlisting}[
    style       =   Go,
]
//In function nestedCall(f): i and callbackToken are int
i := callbackToken  // Line 38: i is of type int
C.callback(unsafe.Pointer(&i))  //Line 45
// In function testCallbackCallers(): Lines 181-194
name := []string{... //save the 12 functions
	"test.nestedCall.func1",  // Line 188: _cgoCheckPointer
	... }
\end{lstlisting}
\captionof{figure}{\myinsert{Test case in callback.go}}
\label{code:cgo:callback}
\end{minipage}
\end{center}

\kw{Validation.} To validate our approach to reducing redundant pointer checks, we use Go's official CGO test cases. The test cases are designed to validate the CGO implementation in the Go compilation toolchain and are executed automatically during the toolchain build process. We pass all tests except one in \codet{misc/cgo/test/callback.go}. This test case is used to verify whether the CGO call chain passes through the 12 functions stored in \codet{name} in sequence (see \autoref{code:cgo:callback} for the core code), where function \kn{test.nestedCall.func1} corresponds to \codet{\_cgoCheckPointer}. However, our approach eliminates an unnecessary pointer check on \codet{unsafe.Pointer(\&i)} in Line 3, so the call chain has one less of 12. In this case, \codet{i} is of type \codet{int} and never holds a pointer, so the elimination is safe and correct. Moreover, we apply our modified Go compiler to 2 existing issues~\citep{issue18184, alertmanager-issue260} and find that our methods can eliminate false positives without requiring developers to modify source code.

\kw{Percentage of unnecessary \codet{\_cgoCheckPointer}.}
We also select 10 most starred repositories and 10 repositories that use the most CGO to demonstrate the effectiveness of our methods. The 10 repositories with the most stars are all listed in \S \ref{sec:empirical:rq1}. The 10 repositories that use the most CGO include \kn{gl, go-gtk, gotch, gotk3, qt, godot-go, go-sdl2, git2go, imagick} and \kn{go-ceph}. They are all repositories that provide bindings.

\input{unnecessary-ptrchk}

We track the original and unnecessary \codet{\_cgoCheckPointer} insertions before and after implementing our new strategy. Out of 50 compiled modules in these repositories, 35 are found to have redundant pointer checks. 
Let \textit{$N_o$} denote the original \codet{\_cgoCheckPointer} count, \textit{$N_u$} denote the detected unnecessary ones via our new strategy, and \textit{$P_u$} represent the percentage of \textit{$N_u$} relative to \textit{$N_o$}.
\autoref{fig:unnecessary-ptrchk} displays the \textit{$P_u$} values for the top 10 modules and the average \textit{$P_u$} across all 50 modules in the 20 repositories, 
where each module name is prefixed with its repository name. %We exclude 2 modules with \textit{$N_o$} and \textit{$N_u$} both equal to 1. Each module name in \autoref{fig:unnecessary-ptrchk} is prefixed with its repository name.}
%\del{\autoref{fig:unnecessary-ptrchk} shows the \textit{$P_u$} value of the 10 modules with the highest \textit{$P_u$} and the average \textit{$P_u$} among all the 50 modules(20 repositories). We exclude 2 modules because their \textit{$N_o$} and \textit{$N_u$} are both 1. Each module name in \autoref{fig:unnecessary-ptrchk} is prefixed with the repository name where the module resides. }
We observe that up to approximately 36.7\% of pointer checks in the \kn{cephfs} module can be safely eliminated. On average across all 20 repositories we can remove 7.2\% of pointer checks, which can significantly enhance the reliability of these codebases. It is noteworthy that most of the top 10 modules are bindings, and their \textit{$P_u$} values are typically higher, especially for \kn{go-ceph}, \kn{git2go}, and \kn{qt}. We delve deeper into why the \textit{$P_u$} values for these three repositories are notably high below.

\subsection{Common scenarios where unnecessary pointer-checks occur}
\label{sec:ptrchk:discuss}

We analyze the initially inserted unnecessary pointer-check instances in the \kn{qt}, \kn{git2go}, and \kn{go-ceph} repositories and observe their commonalities. \cjb{This process followed the open card-sorting method described in \S \ref{sec:qualitative}. Specifically, two authors independently inspected each instance of an unnecessary pointer check, examining the surrounding code context to group similar cases. They then discussed and consolidated their findings to establish a final set of representative scenarios.} Most of \remv{them}\cjb{the scenarios} occur in calls from Go to \kn{libc} memory management or built-in CGO functions like \codet{free}, \codet{memcpy} and \codet{GoBytes}. We summarize contexts of these unnecessary pointer checks as \kn{C1}$\sim$\kn{C3} in \autoref{code:ptrchk:git2go}:

\begin{center}
\begin{minipage}[c]{0.75\textwidth}
%\vspace{-0.1cm}
\begin{lstlisting}[
    style       =   Go,
]
cstart := C.CString(start)
defer C.free(unsafe.Pointer(cstart)) // C1
C.memcpy(unsafe.Pointer(x),unsafe.Pointer(y),size) // C2
C.GoBytes(unsafe.Pointer(x), size) // C3
\end{lstlisting}
\captionof{figure}{\myinsert{Typical contexts of unnecessary pointer checks}}
\label{code:ptrchk:git2go}
\end{minipage}
\end{center}
    
    %\codet{\textbf{Free after CString.}} 
    \mysect{\codet{free} after \codet{CString}}.
    Unnecessary \codet{\_cgoCheckPointer} calls are prevalent in function \codet{free} arguments. These \codet{free} calls consistently follow \codet{CString} within defer statements (C1). \codet{CString} converts Go strings to C char arrays, allocating memory on C heap and returning the starting address. Since Go's garbage collector doesn't manage C memory, manual freeing is necessary. Thus, a deferred \codet{free} function call always follows a \codet{CString}. The \codet{free} function's parameter type is \codet{void *} in C, treated as \codet{unsafe.Pointer} in Go. Consequently, the original Go compiler inserts a \codet{\_cgoCheckPointer} for each call to \codet{free}, leading to numerous unnecessary checks. 

    \mysect{\codet{memcpy}}. In \kn{git2go}, unnecessary \codet{\_cgoCheckPointer} calls are found in \codet{memcpy} invocations (C2). \codet{memcpy} copies bytes from source to destination. Both the first and second parameters of \codet{memcpy} are of type \codet{void *}, resulting in two \codet{\_cgoCheckPointer} checks being inserted regardless of their actual targets.
    
    \mysect{\codet{GoBytes}}. Some unnecessary \codet{\_cgoCheckPointer} checks in \kn{qt} and \kn{git2go} are located in function \codet{GoBytes} calls (C3). \codet{GoBytes} is a built-in Go function and converts C data with explicit length to a Go slice. The first parameter type is \codet{unsafe.Pointer}. Therefore, a \codet{\_cgoCheckPointer} is inserted by \kn{cmd/cgo} regardless of the actual pointer target.
    %\item 

Based on the above findings, it is apparent that unnecessary pointer checks are frequently found in memory-manipulating functions like \codet{free}, \codet{memcpy} and \codet{GoBytes}. We believe there are primarily two reasons: different ways of memory management and memory layouts between Go and C. Go uses GC for memory management, unlike C's manual approach. They also have distinct core data structures, like strings, requiring memory-manipulating functions for compatibility.

\finding{5 to RQ4}{The conservative pointer check strategy in the Go toolchain leads to critical false-positive runtime panics. Permanently resolving such issues requires leveraging the compiler's type inference and pointer analysis passes, which we have submitted as a proposal that has been grouped with an accepted proposal and may see progress in the future. Before the Go team undertakes substantial compiler refactoring, our proposed temporary approach, which focuses on common cases, can also effectively reduce the likelihood of false-positive runtime panics. Additionally, most unnecessary pointer-check instances occur in calls from Go to libc or builtin CGO functions, highlighting an area where developers need to exercise particular caution. }

%% file: unnecessary-ptrchk.tex
\begin{figure}[ht]
    \centering
    \includegraphics[width=0.8\linewidth]{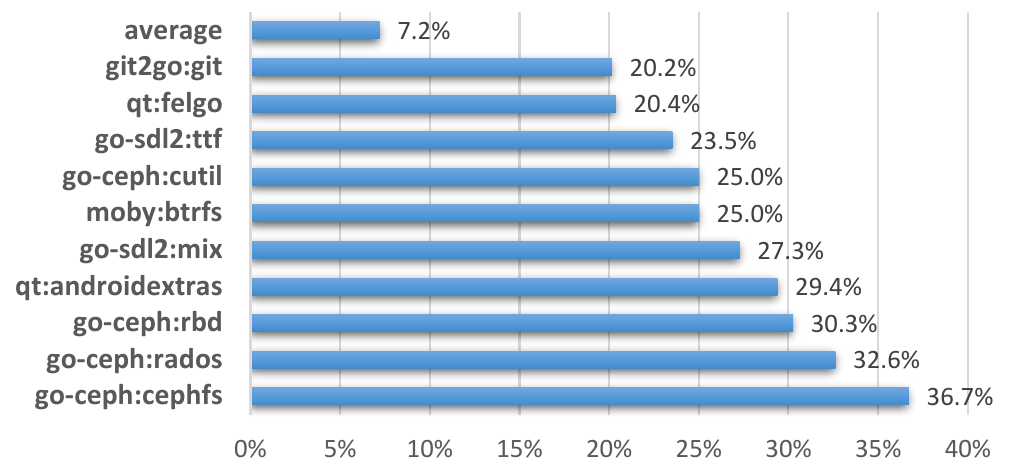}
    \caption{$P_u$ for the top 10 modules and average $P_u$ across all 50 modules (20 repositories)}
    \label{fig:unnecessary-ptrchk}
\end{figure}

%% file: proposal.tex
\section{Implications}
\label{sec:proposal}
%\vspace{-0.3cm}
In this section, we provide practical implications for developers and the Go team based on the findings above.

\subsection{For Developers}
%\vspace{-0.2cm}
\kw{1) Use CGO in suitable contexts to maximize C library benefits.}
%\myins{Based on our findings from RQ1, CGO performs well in accessing system information, system control, and implementing specific functionalities not available in Go. Simultaneously, C libraries generally provide better performance. We recommend that Go developers consider using C libraries through CGO at appropriate times, especially if existing Go libraries cannot meet functionality or performance requirements.}
Our RQ1 findings show that CGO excels in accessing system information, control, and functionalities not available in Go, with C libraries offering better performance.
Go developers can selectively use CGO when Go libraries fail to meet functional or performance needs.

\kw{2) Leverage patterns from mature projects to boost efficiency.}
Common CGO usage patterns identified in RQ2 from well-recognized repositories  (\S \ref{sec:empirical:rq2}) can help developers enhance efficiency. For example, \kn{Unsafe} and \kn{TypCast} involve using the \codet{unsafe} package to accelerate type conversions between Go and C, and \kn{Productivity} involve using wrappers or embedded C code to improve productivity. %But developers should follow the restrictions on passing pointers between Go and C~\cite{cgopointerpass} and be aware of the potential non-portability and incompatibility~\cite{go1compat} of introducing \kn{unsafe} before using. By adopting these practices, developers can reduce the likelihood of errors and improve code maintainability, allowing for better integration between Go and C. Thus, 
We encourage developers to adopts these patterns for more efficient and reliable development.

\kw{\myins{3) Be aware of the potential risks before introducing CGO}.}
Findings in RQ3 (\S \ref{sec:bug}) indicate that CGO can introduce risks, increasing project instability, particularly in build issues (\kn{Build}) and CGO-related bugs (\kn{Bug}). Developers need to be aware of these common CGO issue categories and take proactive measures to enhance project stability. For example, to avoid \kn{Build} issues, developers must address C library build problems across diverse environments. % and must prioritize code quality improvements and thorough testing of CGO-related code. 
These complexities highlight that CGO is not a panacea. Developers should carefully weigh pros and cons before incorporating CGO, and implement safeguards, especially against \kn{Build} and \kn{Bug} issue categories.
%\myins{While CGO offers numerous advantages as mentioned earlier, it also comes with inherent drawbacks. As we found in RQ3, CGO can introduce instability to a project, especially bringing issues in compilation and build processes, as well as CGO-related bugs. Developers must tackle the compilation and build issues of C libraries across various environments (such as different operating systems, hardware, etc.). Regarding CGO-related bugs, developers need to consistently enhance code quality and conduct comprehensive testing on CGO-related code to minimize such issues as much as possible. These challenges present significant difficulties for developers and also remind us that CGO is not a ``magic bullet''. Developers should carefully weigh the pros and cons before introducing CGO into their projects. }

\kw{\myins{4) Use CGO effectively in appropriate methods}.}
\myins{For projects employing CGO, establishing proper usage is critical. Our findings lead us to offer the following recommendations:}

\begin{figure}[htb]
\centering
\begin{minipage}{.45\textwidth}
\begin{lstlisting}[style=Go]
//C1, temporary pointer
p := &b[0]//b is a byte slice
C.test(unsafe.Pointer(p))
//C2, temporary pointer
tmp := unsafe.Pointer(&b[0])
C.test(tmp)

\end{lstlisting}
\end{minipage}
\hfill
\begin{minipage}{.45\textwidth}
\begin{lstlisting}[style=Go, firstnumber=last]
//C3, extra parentheses
C.test((unsafe.Pointer(&b[0])))
//C4 safer style
C.test(unsafe.Pointer(&b[0]))
\end{lstlisting}
\end{minipage}
\caption{Unsafe and safe style}
\label{code:advice:unsafe}
\end{figure}

%Unfortunately, according to our finding and discussion in RQ3 and RQ4, due to the conservative approach Go takes, sometimes false positive runtime panics still occur even when the restrictions are respected. %\del{If developers want to reduce the risk of runtime panic }
%To mitigate this risk without modifying the current Go compiler, developers can convert their coding style from C1$\sim$3 to C4 in \autoref{code:advice:unsafe}. 
\mysect{Change Coding style.} Findings and discussions in RQ4 (\S \ref{sec:ptrchk:issue}) indicate that false positive runtime panics can occur when \codet{unsafe.Pointer} is ued in CGO calls. Until our proposal is substantively advanced and the issue is fully resolved, developers can adjust their coding styles from C1–C3 to C4 in \autoref{code:advice:unsafe} to minimize the occurrence of false positive panics. 
Current Go toolchain can recognize such simplest cases shown in C4 without needing to check the entire \kn{mspan}. The key principle in C4 is avoiding extra parentheses and temporary variables when using CGO combined with \codet{unsafe.Pointer}. %Although the advice cannot solve all problems, it can improve reliability and reduce risk in some common scenarios. 

Another method to eliminate the use of \codet{unsafe.Pointer} is detailed in \S \ref{sec:ptrchk:sol}. Developers writing C code for CGO programs are advised not to use \codet{void *} as the C function parameter type when the type can be determined. By specifying a concrete pointer type (e.g., \codet{int *}) as the parameter type for the C function, there is no need to convert the pointer to \codet{unsafe.Pointer} when calling this C function in Go, thus avoiding unnecessary pointer checks.

\mysect{Use Our Temporary but Effective Approach}
If developers do not want to modify their existing code, they can use the temporary but effective approach we propose in \S \ref{sec:ptrchk:sol}. Our approach is based on Go 1.17.7 and has already been open-sourced on \url{https://github.com/S4Plus/CGO-ARTIFACT}. Although this approach cannot guarantee the elimination of all unnecessary pointer checks, it handles common cases. These adjustments can help developers mitigate the occurrence of such false-positive panics to some extent, without requiring significant changes to their codebase. 

\subsection{For the Go team}
%\vspace{-0.1cm}
%\del{According to our findings, there are gaps between the definition of the constraints on passing pointers across CGO and the current Go implementation used for enforcing these constraints. The existing Go implementation produces some false positives. Although the Go team has used certain tricks to mitigate them in some common scenarios (such as \autoref{code:advice:safe}), false positives can still occur in other scenarios (such as \autoref{code:advice:unsafe}).}
Our findings show discrepancies between defined constraints on passing pointers across CGO and the current Go implementation for enforcing the constraints, resulting in false positive panics. Although the Go team has mitigated common scenarios (C4 in \autoref{code:advice:unsafe}), false positives persist (C1$\sim$C3). Based on our analysis, we recommend the following actions:%Our straightforward improvements underscore there is still substantial room for enhancement in this area. 

\kw{1) Push forward with the accepted proposal.} 
Given that we have submitted a proposal for a permanent fix, we encourage the Go team to actively push forward with the accepted proposal and complete the compiler refactoring as soon as possible. This will significantly improve the reliability of the Go compilation toolchain, positively impacting both the Go development community and the broader software ecosystem. 

\kw{2) Give clear documents for potential risks.} 
% 如果上述proposal在短期无法推进，我们建议Go team在CGO相关的documents中强调当前在restriton 和 Go toolchain 之间的discrepancies，并提供一些官方的开发建议（比如上面的C1到C4的coding style），以最大限度降低开发门槛，减少开发误解
If the aforementioned proposal cannot be implemented in the short term, we recommend the Go team explicitly document the current discrepancies between CGO restrictions and the Go toolchain. The documentation should also include official development guidelines, such as the coding styles described in C1 to C4 to lower the development barrier and minimize misunderstandings. Furthermore, we also recommend including practical examples, best practices, known edge cases and avoidable pitfalls in the documentation to help developers write more reliable CGO code, enabling them to address issues earlier rather than encountering them unexpectedly at runtime and improving code reliability.

%% file: threats.tex
\section{Threats to Validity}
\label{sec:threats}

\cjb{This section outlines potential threats to the validity of our findings, categorized as internal and external threats, and details the mitigation strategies we adopted.}

\mysect{\remv{Internal Threats.}} 
\subsection{\cjb{Internal Threats}}
\cjb{\mysect{Sampling Bias.}}
\remv{The first internal threat is our sampling bias. Our study concentrates on popular open-source Go projects, potentially missing less-known or private projects that might exhibit distinct CGO usage patterns or issues. Besides, we only identify CGO-related problems from GitHub issues containing specific keywords, which may result in overlooking problems from other sources. This could lead to underestimating the actual severity of the problems related to CGO.}\cjb{Our focus on popular open-source projects and keyword-based issue searches may overlook patterns in less-known projects and underestimate CGO's challenges. To mitigate this, we clearly define our selection criteria, which deliberately scopes the study to active, popular projects. This ensures our findings are grounded in influential community practices and offer relevant insights for contemporary Go developers.}

\cjb{\mysect{Accuracy of Our Analysis Tools.}}
\remv{What's more, \mbox{\mysys} may suffer from false negatives or false positives, potentially affecting the accuracy of the reported statistics.}\cjb{A general, inherent limitation of static AST analysis is its inability to detect dynamic features. As \mysys relies on this technique, it may miss CGO calls generated dynamically or expanded from complex C macros, potentially leading to false negatives. To mitigate this, we built \mysys atop Go's official \kn{go/parser} package, which robustly handles the complexities of both Go and CGO code, ensuring broad applicability across projects. While dynamic analysis could capture such calls, it introduces other trade-offs and is beyond the scope of our large-scale static study.}
%\cjb{\mysys{} relies on static analysis via AST parsing and it may fail to detect CGO calls generated dynamically or expanded from macros on the C side, cause false negatives and potentially affect the accuracy of the reported statistics. In order to improve the accuracy as much as possible, \mysys{} is built upon Go's official and robust \kn{go/parser} package, which supports parsing both Go and CGO code, so as to be applicable to most projects.}

\cjb{\mysect{Subjectivity in Qualitative Analysis.}}
\cjbr{Furthermore, the}{The} interpretation of issues such as compilation errors, bugs, and runtime crashes may be influenced by researchers' preconceived notions or biases, leading to subjective assessments. \cjb{In order to minimize the impact of these subjective factors, we use a multi-step qualitative analysis method based on open card sorting and Cohen-Kappa metric as described in \S \ref{sec:qualitative}.}

\mysect{\remv{External Threats.}} 
\subsection{\cjb{External Threats}}
The first external threat concerns generalizability. Our findings pertain specifically to popular open-source Go projects on GitHub and may not extend to other Go projects hosted on different platforms, private projects, or those from specialized domains with fewer stars. 
Secondly, the study's context within Go projects may overlook unique challenges or usage patterns in projects using different development methodologies, limiting broader applicability. Lastly, changes in Go language features, compiler optimizations, or CGO-related practices may impact the relevance and applicability of the insights over time. 

%% file: conclusion.tex
\section{Conclusions}
\label{sec:conclusion}
Our empirical study reveals that CGO currently has an adoption rate of 11.3\% in the community, which is a significant portion that cannot be ignored. Moreover, the usage of CGO is typically concentrated in a few specific files or modules. Using CGO can enhance the flexibility and scalability of Go projects, and developers often use it to interact with the system and achieve better performance. Common patterns related to CGO in popular Go projects include communication between Go and C, type conversion, and performance optimizations. Developers can refer to these well-recognized patterns and purposes.

Moerover, we also find that while using CGO has many advantages, it also introduces some issues, such as those related to project building. Among the issues related to CGO, pointer check-induced false-positive panics are particularly critical. This critical issue arises from the limitations of the current Go compilation toolchain, as it cannot determine the type of the object pointed to by \codet{unsafe.Pointer} during CGO preprocessing. To address this, adopting a temporary strategy that focuses on common cases can effectively improve the situation. A permanent fix requires refactoring the Go compilation toolchain to make CGO a part of the compiler and reuse results from the compiler’s type inference and pointer analysis passes. A proposal for this permanent fix has already been submitted and grouped with an accepted proposal, and we expect the Go team to make further progress and implement the fix in the future. 

Finally, we provide developers with some implications on how to efficiently and safely use CGO to enhance project development efficiency and improve safety based on our findings. We also encourage the Go team to push forward with the proposal as soon as possible, because by enhancing the toolchain’s stability, developers will be able to build more robust applications.

%% file: data.tex
\section*{Data Availability}
We make our data and code publicly available at \url{https://github.com/S4Plus/CGO-ARTIFACT}

%% file: ack.tex
\section*{Acknowledgements}
This work was supported by the National Natural Science Foundation of China [grant numbers 62272434].